\documentclass[twocolumn]{aastex631}
\usepackage{makecell}
\usepackage{amsmath}
\usepackage{mathtools}

\defcitealias{Planck2016}{Planck Collaboration et al. 2016}

\begin{document}

\title{Heavy Seeds and the Multi-scale Fueling of Early Supermassive Black Holes and Little Red Dots}

\author[0000-0002-6038-5016]{Junehyoung Jeon}
\affiliation{Department of Astronomy, University of Texas, Austin, TX 78712, USA}
\affiliation{Cosmic Frontier Center, The University of Texas, Austin, TX 78712, USA}
\author[0000-0003-0212-2979]{Volker Bromm}
\affiliation{Department of Astronomy, University of Texas, Austin, TX 78712, USA}
\affiliation{Cosmic Frontier Center, The University of Texas, Austin, TX 78712, USA}
\affiliation{Weinberg Institute for Theoretical Physics, University of Texas, Austin, TX 78712, USA}
\author[0000-0002-4966-7450]{Boyuan Liu}
\affiliation{Institut für Theoretische Astrophysik, Zentrum für Astronomie, Universität Heidelberg, D-69120 Heidelberg, Germany}
\author[0000-0003-1252-9916]{Stella S. R. Offner}
\affiliation{Department of Astronomy, University of Texas, Austin, TX 78712, USA}
\author[0000-0003-1282-7454]{Anthony J. Taylor}
\affiliation{Department of Astronomy, University of Texas, Austin, TX 78712, USA}
\affiliation{Cosmic Frontier Center, The University of Texas, Austin, TX 78712, USA}
\author[0000-0001-8519-1130]{Steven L.~Finkelstein}
\affiliation{Department of Astronomy, University of Texas, Austin, TX 78712, USA}
\affiliation{Cosmic Frontier Center, The University of Texas, Austin, TX 78712, USA}

\email{junehyoungjeon@utexas.edu}

\begin{abstract}
The formation of supermassive black holes (SMBHs) in the early Universe has been a longstanding mystery due to the brief period available to assemble them. More recently, the  unexpected discovery with the James Webb Space Telescope (JWST) of Little Red Dots (LRDs), compact sources exhibiting signatures of active galactic nuclei (AGN), has accentuated this challenge. Cosmological simulations investigating their formation and evolution confront the limitation that the compact scale of SMBH/LRD physics cannot be simultaneously resolved with the larger-scale cosmological environment. We address this limitation by utilizing a cutout technique, selecting a targeted region around an SMBH seed from a cosmological parent simulation and re-running the cutout at higher (sub-pc) resolution for shorter periods. We explore two SMBH formation scenarios: the standard, light seed stellar-remnant case, and the heavy-seed, direct collapse black hole (DCBH) channel. Within the high-resolution achieved by the cutouts, the DCBH, due to its larger mass, is able to induce a buildup of high density gas nearby and accrete gas in bursts of super-Eddington phases, growing in mass and producing an LRD-like spectrum. In contrast, the stellar remnant cannot do so and remains dormant, unable to affect the nearby gas or significantly evolve by gas accretion alone. Therefore, we find that observed high-redshift SMBHs and LRDs require the existence of heavy seeds to assemble dense gas around themselves, enabling efficient growth and producing red optical spectra with strong Balmer breaks resembling LRDs.


\end{abstract}


\section{Introduction} \label{sec:intro}

How do supermassive black holes (SMBHs) and Little Red Dots (LRDs) emerge in the early Universe \citep[e.g.,][]{Smith2019_2,Woods2019,Inayoshi2020,Matthee2023,Kokorev2024_lrd,Kocevski2023,Akins2024}? This question has further risen in prominence with the recent discoveries from the James Webb Space Telescope (JWST). SMBHs/active galactic nuclei (AGN) have been found in larger numbers and at higher redshifts than previously expected \citep{Bogdan2023,Maiolino2023,Napolitano2024,Taylor2025,Chavez2025,Kocevski2023,Onoue2023,Furtak2023,Greene2023,Taylor2024}, along with the revelation of the mysterious LRDs, exhibiting both AGN (e.g., broad spectral lines) and non-AGN like (e.g., lack of X-ray emission and variability) characteristics \citep{Akins2024,Kocevski2023,Liu2026}. 

Various theoretical models have been proposed to explain high-redshift SMBHs and LRDs, some overlapping. To produce quasars with their large masses ($\gtrsim10^8$ M$_\odot$) at $z\gtrsim6$ \citep{Wu2015,Banados2018,Zubovas2021,Fan2023}, extreme conditions have to be invoked. One possibility is that the initial seed black holes (BHs) formed with somewhat higher masses, compared to locally observed stellar remnants at few solar masses. The first, or Population~III (Pop~III) stars, born out of metal-free primordial clouds, are expected to have a top-heavy initial mass function \citep[IMF;][]{Stacy2016,Hirano2017,Latif2022}, and produce more massive BH seeds around $\sim100$ M$_\odot$ \citep{Madau2001,Heger2003,Johnson2007_2}. After formation, these light seeds will grow through gas accretion and BH mergers \citep{Jeon2012,SmithRegan2018,Bhowmick2024_2,Trinca2022,Porras2025}, requiring super-Eddington accretion to reach the observed high-$z$ SMBH masses \citep{Jeon2025}. Interestingly, such conditions of super-Eddington accretion could suppress X-ray emission/variability, in line with what is observed for LRDs \citep{Madau2026,LiuH2025,Secunda2026,Pacucci2024}. 

Alternatively, to mitigate the need for sustained episodes of efficient accretion \citep{Haiman2001}, heavy seed channels have been proposed, where initial seed masses already reach $\sim$10$^5$\,M$_\odot$ \citep[e.g.,][]{Becerra2018b}. Direct Collapse Black Holes (DCBHs) constitute a well-established model of heavy seed formation -- forming from the collapse of a massive gas cloud, with a supermassive star (SMS) and/or quasi-star (QS; \citealt{Begelman2006}) as intermediate, short-lived stages. The gas cloud must be extremely metal-poor \citep[$Z\lesssim 10^{-3}\ \rm Z_\odot$,][]{Chon2024} and requires rare conditions, prohibiting gas cooling to low temperatures to allow cloud-collapse without vigorous fragmentation \citep[e.g.,][]{Bromm2003,Lodato2006,Lodato2007,Johnson2013,Wise2019,Haemmerl2018,Haemmerle2020,Luo2020}. Given the conditions required for their formation, DCBHs are expected to reside in low stellar-mass regions surrounded by a dense gas envelope, possibly producing LRDs \citep{Cenci2025,Jeon2025_lrd,Pacucci2026,Whalen2026}. Similarly, the transitory SMS and QS phases have been proposed to be LRDs \citep{Chisholm2026,Kokorev2026_sms,Begelman2025,Santarelli2026}.\footnote{Other heavy seed scenarios exist, such as BH formation from runaway collisions of (proto-)stars in dense stellar clusters with high gas inflow rates \citep{Zwick2023,Reinoso2023,Klessen2023,Gaete2024}, or primordial BHs (PBHs) formed from Big Bang overdensities \citep{Dayal2024,Zhang2025,Zhang2026}.\label{footnote}}

Still, even for heavy seeds, efficient gas accretion, possibly even super-Eddington accretion, will be required to produce the observed early SMBHs with masses orders of magnitude above the proposed seed masses \citep[e.g.,][]{Jeon2024,Jeon2025,Jeon2025_lrd,Jeon2026_ff}. BH mergers could facilitate BH mass growth at early times, but gas accretion will still be necessary later on \citep{Bhowmick2026_acc}. 
However, self-consistently modeling BH growth into SMBHs is challenging due to the difference in scales covering their evolution. The BH accretion physics operates at sub-pc scales, whereas the cosmological environment providing the baryonic material for accretion evolves on larger kpc to Mpc scales. Cosmological simulations focused on galaxy-scale evolution cannot simultaneously resolve the small-scale BH physics, and thus require approximate subgrid models to represent BH evolution \citep[e.g.,][]{Jeon2023,SmithRegan2018,Liu2020,Volonteri2016}.  

For LRDs, an emerging model proposes that they host an accreting central object, likely an SMBH, surrounded by a dense, dust-free gas cocoon \citep{Inayoshi2025_bl}. The dense gas produces a red spectrum with a strong Balmer break and exponential profiles, well-matching the observed LRD spectra \citep{deGraff2025,Kokorev2025,Taylor2025,Naidu_BHstar2026}. The required hydrogen densities are extreme $(10^8-10^{10}$ cm$^{-3})$ and a massive reservoir of gas will be needed to sustain the dense gas envelope as the SMBH accretes \citep{Inayoshi2025_lrd,Jeon2026_ff}. Again, similar to the question of BH accretion, cosmological simulations cannot resolve the scale of proposed LRD configurations and cannot confirm the existence of the proposed dense gas cocoons around SMBHs. Instead, a simulation that covers the large-scale cosmological gas inflow to the small-scale BH environment will be required.

To address these questions of BH seed growth and the nature of LRDs, we perform high-resolution cosmological simulations in this work. Specifically, we utilize a cutout technique, where we excise a targeted region from a large-volume cosmological simulation, evolving it for short periods at a higher resolution \citep{Clark2011,Stacy2013}. We select regions near BH seeds that formed under different formation channels in the simulation, so that both the gas accretion and structure can be followed at extremely high, sub-pc resolution, elucidating the multi-scale nature of early Universe BH evolution and the emergence of LRDs, from the cosmic web to near-BH scales. 
Other works have also tried to resolve both the cosmological and near-BH scales simultaneously \citep[e.g.,][]{Hopkins2023,Hopkins2024,Hopkins2024_2}, but at a high computational cost. Our numerical approach, involving relatively low particle numbers, is computationally much less expensive, thus enabling a broader exploration of parameter space. 

In this work, we use \textit{Planck} cosmological parameters \citepalias{Planck2016}: $\Omega_m = 0.315$, $\Omega_b = 0.048$, $\sigma_8 = 0.829$, $n_s = 0.966$, and $h = 0.6774$. We explain our simulation models and cutout setup in Sec.~\ref{sec:method}, present our results in Sec.~\ref{sec:results}, discuss the implications for SMBH and LRD assembly in Sec.~\ref{sec:discussion}, and summarize the work in Sec.~\ref{sec:conclude}.

\begin{deluxetable*}{cccccccc}[htb!]
\tabletypesize{\footnotesize}
\tablenum{1}
\tablecolumns{1}
\tablecaption{Summary of the simulation setup\label{tab:sims}}
\tablewidth{0pt}
\tablehead{
\colhead{Simulation} & \colhead{\makecell{Linear Size}} & \colhead{\makecell{DM \\ particle mass}} &\colhead{\makecell{Gas \\ particle mass}} & \colhead{\makecell{Gas minimum gravitational \\ softening length}} \\ \colhead{} &\colhead{}& \colhead{M$_\odot$} &\colhead{M$_\odot$} & \colhead{} 
}
\startdata
Parent  &2 cMpc$/h$  &$8\times10^3$& N/A & N/A \\
Zoom &0.45 cMpc$/h$ & 800 & 150& 10 cpc$/h$\\
Cutout &200 pc &800& 1.5& 0.1 pc  \\
Cutout\_Highres &200 pc &800& 0.3& 0.02 pc\\
\enddata
\centering
\end{deluxetable*}

\begin{deluxetable*}{cccccccc}[htb!]
\tabletypesize{\footnotesize}
\tablenum{2}
\tablecolumns{1}
\tablecaption{Cutout simulation suite\label{tab:cutouts}}
\tablewidth{0pt}
\tablehead{
\colhead{Cutout} & \colhead{\makecell{Initial Redshift}} & \colhead{\makecell{Gas Particle\\ Number}} & \colhead{\makecell{Total Gas\\ Mass}} & \colhead{\makecell{Initial BH\\ Mass}} &\colhead{\makecell{BH Accretion}} & \colhead{\makecell{BH Feedback}} \\ \colhead{} &\colhead{} &\colhead{}&\colhead{M$_\odot$}& \colhead{M$_\odot$} &\colhead{} & \colhead{} 
}
\startdata
Stellar&  10.1&2.56$\times10^{5}$ &3.75$\times10^{5}$ &140& No & No \\
DCBH& 9.1& $7.3\times10^5$ &1.1$\times10^6$ & $1.5\times10^5$ & No& No\\
DCBH\_Accrete&9.1& $7.3\times10^5$ &1.1$\times10^6$ & $1.5\times10^5$ & Yes& No  \\
Stellar\_Accrete&10.1& 2.56$\times10^{5}$&3.75$\times10^{5}$ &140& Yes& No  \\
DCBH\_Feedback &9.1& $7.3\times10^5$ &1.1$\times10^6$ &  $1.5\times10^5$ &Yes& Yes\\
DCBH\_Highres &9.1& 3.7$\times10^6$ &1.1$\times10^6$ & $1.5\times10^5$ & No& No\\
\enddata
\centering
\end{deluxetable*}

\begin{figure*}[!htb]
    \centering
    \includegraphics[width=0.9\textwidth]{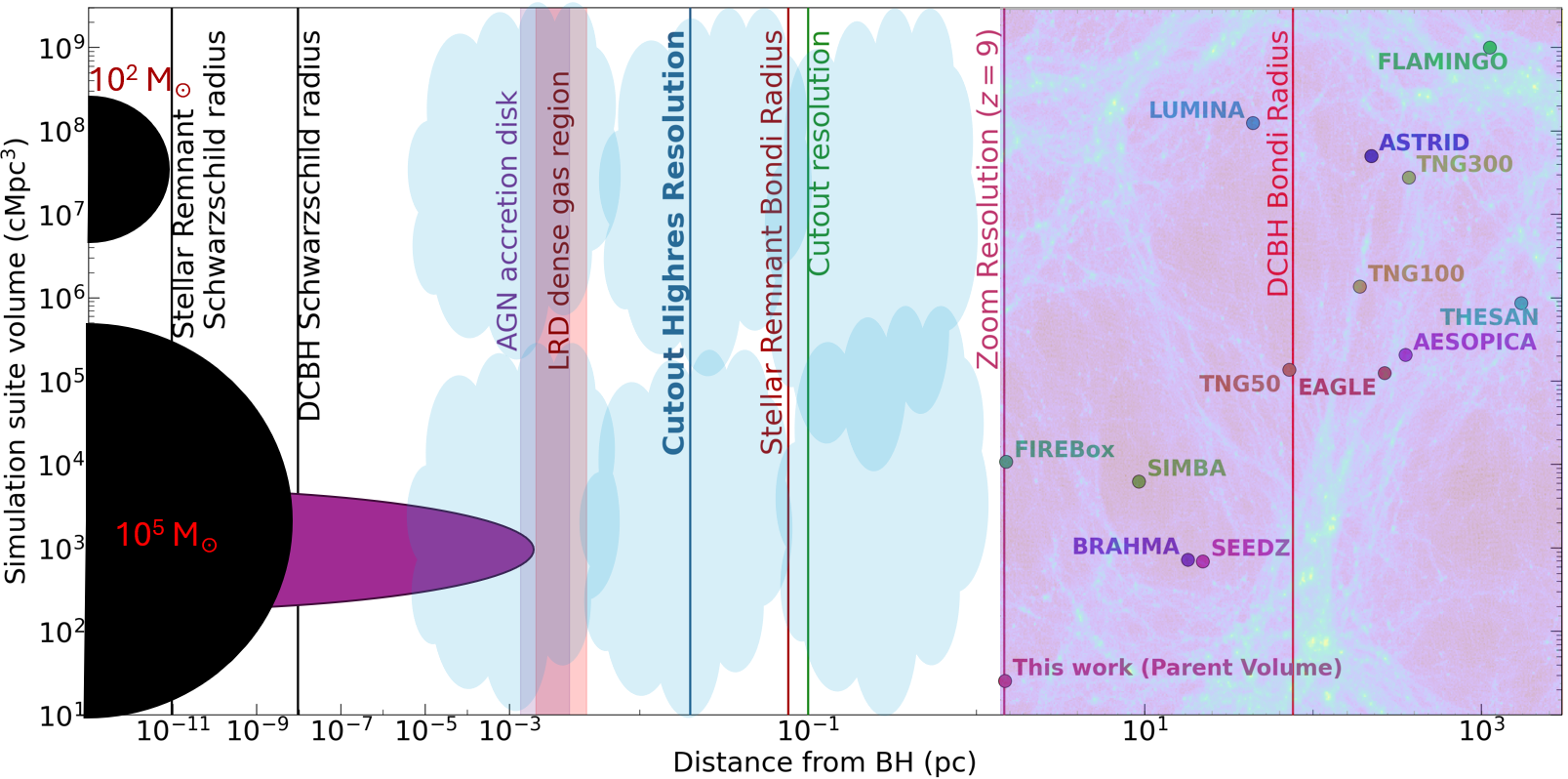}
    \vspace{-0.8em}
    \caption{Key spatial scales for BH evolution. We indicate the inferred size of the high-density regions in LRDs \citep{Inayoshi2025_lrd,Kokorev2025}, the AGN accretion disk \citep{Guo2022}, as well as the Schwarzschild and Bondi radii (Eq.~\ref{eq:br}) for a 100 M$_\odot$ stellar remnant BH and $10^5$ M$_\odot$ DCBH, assuming average gas temperature of $10^3$ K (see Fig.~\ref{fig:Tdist}). We also show the spatial resolutions (scaled to $z=9$) and volumes achieved by other select simulations \citep{Crain2015,Pillepich2019,Zier2026,Dave2019,Schaye2015,Feldmann2023,Kannan2025,Zhou2026_astrid,Bhowmick2024_2,Koudmani2026,Prole2026_seedz}, compared to this work (Table~\ref{tab:sims}). All cosmological simulations have resolutions orders of magnitude above relevant BH scales, and thus cannot directly capture the detailed BH physics. The exploratory simulation in this work, while smaller in volume compared to other studies, is used to test the cutout method that can reach much closer to the scale of BH accretion and resolve the Bondi radius. We note that even the highest resolution obtained in the cosmological simulations is many orders-of-magnitude larger than the general-relativistic BH event horizon, but that resolving the Bondi radius is achievable.  
        }
    \label{fig:enclosed}
\end{figure*} 

\vspace{-4em}
\section{Numerical Methodology} \label{sec:method}

We use the \textsc{gizmo} code \citep{Hopkins2015}, employing the Lagrangian meshless finite mass (MFM) hydro solver, and the gravity solver inherited from \textsc{gadget-2} \citep{Springel2005}. We specifically use the modified version of \textsc{gizmo} \citep{Liu2020,Jeon2023,Jeon2024} with distinct sub-grid models for star formation (SF) and feedback; BH formation, accretion, and feedback; as well as primordial chemistry, cooling, and metal enrichment \citep{Jaacks2018,Jaacks2019}. 

\subsection{Simulation setup}\label{sec:simset}

\begin{figure*}[!htb]
    \centering
    \includegraphics[width=0.9\textwidth]{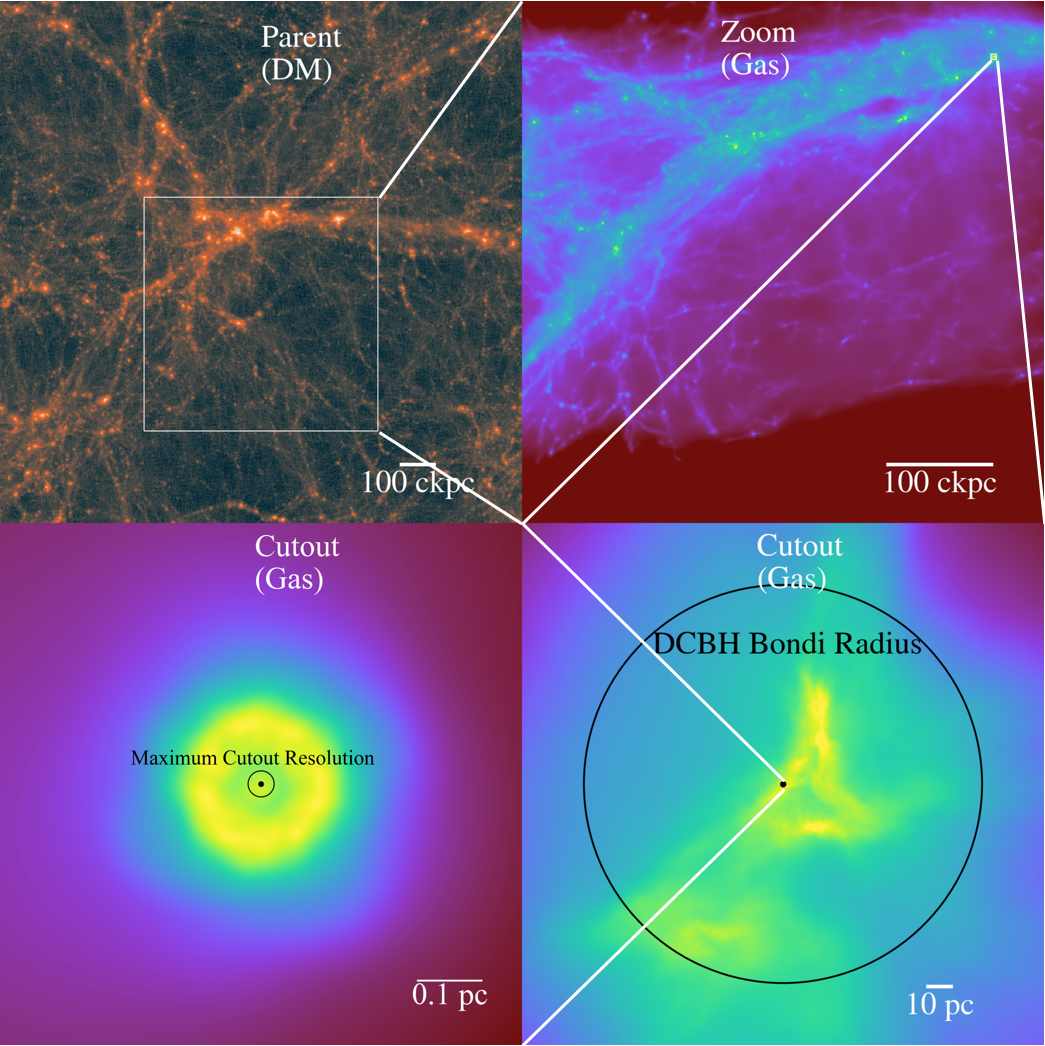}
    \caption{Overview of the nested set of simulations performed in this work (see Table~\ref{tab:sims}), illustrating the large dynamical range in spatial scales and mass achieved. The parent simulation shows the DM distribution, while the other panels show the gas distribution. We indicate the approximate location in the respective larger-scale boxes for the subsequent refinement. The cutout simulation panels (bottom row) mark the location of the DCBH seed with a black dot. Using the cutout technique, we are able to simulate the process of SMBH assembly from cosmological scales down to sub-pc scales (indicated with the black circle in the bottom-left panel), thus resolving both the large-scale gas inflow and the small-scale behavior near the accreting BH, including the formation of the gaseous disk around the BH particle, a natural consequence of the angular momentum of the infalling gas. We note that the cutouts can resolve the Bondi scale (indicated with the black circle in the bottom-right panel).}
    \label{fig:overview}
\end{figure*}

The simulation initial conditions are generated with the \textsc{MUSIC} code \citep{Hahn2011} at $z=99$. We first run a parent simulation of 2 cMpc$/h$ in linear size with only dark matter (DM) particles with masses $8\times10^3$ M$_\odot$. Using \textsc{Rockstar} \citep{Behroozi2013} in post-processing to identify DM halos, we select a $\sim7\times10^8$ M$_\odot$ halo at $z\simeq6$. We then identify a region that contains all the DM particles of the selected halo as the zoom-in region. 

The zoom-in simulation has a linear size of 450 ckpc$/h$, includes baryons, and has an increased resolution with DM particle masses of $\sim800$ M$_\odot$ and baryon particle masses of $\sim150$ M$_\odot$. The high-resolution zooms are run to $z\sim4-5$ to examine the formation of SMBHs. 

In the zoom-in simulation, we calculate the virial parameter $\alpha$ of the gas particles within 100 pc (in physical units) of each BH particle that formed across redshifts, 
\begin{equation}\label{eq:virial}
    \alpha = \frac{2E_{\rm th}}{E_{\rm pot}}\simeq\frac{c_{\rm s}^2R_{\rm eff}}{G(M_{\rm gas}+M_{\rm DM})}
\end{equation} 
where $E_{\rm th,pot}$ is the thermal/gravitational potential energy of the cutout system, $c_{\rm s}$ the gas sound speed, $R_{\rm eff}$ the weighted average of the gas and DM particle distance from the cutout center (BH location), $M_{\rm gas}$ the mass of the gas, and $M_{\rm DM}$ the mass of DM particles in the region. Lower values of $\alpha$ suggest a more bound gas system, and we choose, for a DCBH or a stellar BH, systems with high mass but also low $\alpha$, to consider strongly bound configurations so that optimistic BH growth can be followed. 

The chosen system is then converted into a cutout containing the gas, DM, and central BH particle. The gas particles within the 100 pc region are split by a factor of 100 \citep[see also][]{Kitsionas2002}, so that they have masses of $\sim1$ M$_\odot$ and $\sim0.1$ pc minimum softening length resolution. The split gas particles are placed randomly within the smoothing kernel of the original gas particle \citep{Clark2011,Stacy2013}. We do not split the DM particles, to not introduce unphysical gravitational structure in the cutout not inherited from the original cosmological simulation. We further perform a run where the gas particles are split by a factor of 500, resulting in mass and length resolutions of $\sim0.2$ M$_\odot$ and $\sim0.02$~pc, respectively. In the cutouts, we disable SF and stellar feedback, as the gas particle masses are now too low for our respective models that have been designed for cosmological simulations with higher gas particle masses. For BH models, specifically accretion and feedback (see Sec.~\ref{sec:bhacc} and \ref{sec:bhfeed}), we perform runs where they are disabled, such that the BH only interacts gravitationally with the gas, and where they are enabled, to assess the effects of high resolution on BH growth and energy output.

The cutouts (see Sec.~\ref{sec:cutset} for details) are run for $\lesssim10$\,Myr. We thus avoid unphysical edge effects, resulting from the artificial outer (vacuum) boundary conditions of the cutout region, where all external material has been removed. We can estimate the period over which this approximation is valid with the sound crossing time to traverse the cutout region. Given an average gas temperature of $\sim10^3-10^4$\,K, corresponding to a sound speed of $\sim10^{-5}$ pc yr$^{-1}$, the distance corresponding to the 10~Myr period is $\sim100$~pc, half the size of the cutout region. We accordingly limit our analysis to $<8$ Myr periods. Table~\ref{tab:sims} summarizes the simulation setup and Figs.~\ref{fig:enclosed} and \ref{fig:overview} illustrate the wide range of length scales involved. We further verify the validity of the cutout methodology with an idealized test case of a collapsing spherical gas cloud, as described in Appendix~\ref{sec:appendix}.

\subsubsection{Cutout selection}\label{sec:cutset}

We run multiple cutout simulations for two select BH seeds while varying the included BH subgrid models (accretion and feedback). We specifically consider one stellar  remnant BH and one DCBH seed, selected from the zoom-in simulation to represent cases where the nearby gas is closest to being self-gravitating, as determined by Eq.~\ref{eq:virial}. 
The DCBH cutout had $\alpha\simeq0.3$ and the Stellar cutout $\alpha\simeq0.2$. 
We note that these values for the virial parameter are lower limits, not including the modest contribution of supersonic, turbulent motions in the velocity dispersion (which would result in $\alpha\sim 1$).
For both seeds, we run cutouts with BH accretion turned off and on. For the DCBH case only, we further perform runs with increased resolution by a factor of 5 (DCBH\_Highres run), and with BH feedback enabled as well (DCBH\_Feedback run). We do not perform corresponding runs for the stellar remnant seed, as we find minimal changes in the stellar BH evolution and mass growth (see Sec.~\ref{sec:results}). In Table~\ref{tab:cutouts}, we summarize the properties of the cutout simulations that have been carried out.

\subsection{Stellar Models}

For the zoom-in runs, we employ the SF and feedback modules developed in \citet[][]{Jaacks2018,Jaacks2019,Liu2018,Liu2020}. We include models for both the first, metal-free Pop~III, as well as metal-enriched Population~II (Pop~II) stars. The Pop~III model has been tested against extremely high-resolution simulations \citep[e.g.][]{Johnson2007,Ritter2012,Ritter2015}, and the Pop~II model calibrated with observed cosmic star formation histories across $z\sim5-15$. Individual stars cannot be resolved with the resolution reached in the zoom-in simulations, such that each stellar particle represents a stellar population, characterized by Pop~III and Pop~II IMFs from \citet{Jaacks2018,Jaacks2019}. 

We further include stellar feedback in the form of Lyman-Werner (LW) radiation, photo-ionization heating, mechanical winds, and supernova (SN) explosions \citep{Ciardi2005}. Sufficiently strong LW flux enables DCBH formation in the simulation by destroying molecular hydrogen to limit gas cooling at low temperatures. If gas collapse to form stars is delayed until the DM halo grows to be more massive than the atomic-cooling threshold $(>10^7$ M$_\odot$), vigorous fragmentation into less massive stars is prevented, and the high rates of gas inflow may instead trigger the formation of massive BH seeds \citep[see also][]{Liu_acc2024}. For LW and photo-ionization radiation, we include both global and local contributions. 
For full details of the SF and feedback models, we refer the reader to \citet{Jaacks2018,Jaacks2019,Liu2020}.

\subsection{Black hole seeding}

As BH seeds, we consider Pop~III stellar remnants and DCBHs. For Pop III remnants, we use the massive seed model of \citet{Liu2020}, where all of the stellar particle mass is converted into the BH seed mass. When a star particle reaches the end of its life after 3~Myr, we assume that the stars of the particle will merge to form a single BH, effectively converting the stellar particle to a BH sink particle. 

For DCBH seeding, we impose multiple conditions to represent the dense, hot, and metal-poor environments necessary for formation \citep[e.g.,][]{Bromm2003,Natarajan2017,Ardaneh2018,Wise2019,Chon2021}. When a gas particle's (hydrogen number) density reaches $n_{\rm DCBH} \geq 2\times10^3$ cm$^{-3}$, has a temperature of $7000<T/\text{K}<10^4$, a metallicity of $Z<Z_{\rm DCBH} = 2\times10^{-4}$ Z$_\odot$, and H$_2$ abundance of $x_{\rm H_2}<10^{-6}$, it is considered as a DCBH candidate. Furthermore, the gas particles in the neighborhood of the DCBH candidate (within the BH kernel) must also satisfy the temperature, metallicity, and H$_2$ abundance criteria, with the DCBH candidate being the densest gas particle among its neighbors. When all these conditions are met, the DCBH candidate is converted to a BH sink particle.

However, as the gas particle masses in the zoom-in simulation ($\sim150$ M$_\odot$; Sec.~\ref{sec:simset} and Table~\ref{tab:sims}) are much lower than the fiducial DCBH mass of $\sim10^5$ M$_\odot$ \citep{Pacucci2017,Becerra2018a,Becerra2018b,Basu2019}, the initial sink particle mass formed is much lower than the predicted DCBH mass. Thus, soon after formation, we force the DCBH sink particle to stochastically swallow nearby gas particles \citep{Springel2005_2,Liu2020}, around 1,000 of them, to reach $\sim10^5$ M$_\odot$ while maintaining mass conservation. To enable such rapid initial growth through aggressive gas swallowing, we use two different mass values to characterize the sink particle. The first is the internal BH mass $M_{\rm BH}$ set at formation to 1,000 times that of the original gas particle mass. This mass is used to determine BH accretion and feedback parameters in the simulation. The second is the dynamical mass $M_{\rm dyn}$ of the sink particle, set at formation to the progenitor gas particle mass. This mass is used to determine the dynamical interactions of the sink particle with other particles in the simulation. $M_{\rm dyn}$ of the sink particle is only updated when the sink particle can swallow a gas particle. Thus, when $M_{\rm BH}$ of the particle is set to be larger than $M_{\rm dyn}$ at formation, the sink particle will try to swallow nearby gas particles to match the two masses, while maintaining mass conservation (see Section~\ref{sec:bhacc}).
The criterion that the environment around the DCBH candidate also satisfies DCBH formation conditions establishes that such sum of gas particles is not unphysical, as it results in the formation of DCBH seeds in high gas density environments, such that the seed can immediately swallow the nearby gas particles to reach DCBH masses. 

\subsection{Black hole accretion and growth} \label{sec:bhacc}

For BH accretion, we use the Bondi-Hoyle equation \citep{Bondi1944}:
\begin{equation}
    \dot{M}_{\rm Bondi}=4\pi\frac{(GM_{\rm BH})^2\rho_{\rm g}}{c_{\rm s}^{3}}
    \label{eq:bondi}
\end{equation}
where $M_{\rm BH}$ is the mass of the BH and $\rho_{\rm g}$ is the density of gas particles in the BH kernel. The kernel size is adaptive to include 64 gas particles nearest to the BH.

When BH accretion is enabled for the cutouts, we instead use the modified Bondi-Hoyle equation accounting for the angular momentum of the gas \citep{Tremmel2017,Liu2020}, as the gas structure around the BH is well resolved. We first calculate the characteristic gas rotational velocity around the BH as $v_{\theta}=j/\epsilon_g$, where $j$ is the specific angular momentum of the nearby gas and $\epsilon_g$ is the gravitational softening length of the BH particle. When compared against the gas bulk velocity, $v_{\rm bulk}$, if $v_{\theta}\leq v_{\rm bulk}$, angular momentum is negligible and we use
\begin{equation}
    \dot{M}_{\rm Bondi}=4\pi\frac{(GM_{\rm BH})^2\rho_{\rm g}}{(c_{\rm s}^2+v_{\rm g}^2)^{3/2}}
    \label{eq:bondi_2}
\end{equation}
where $v_{\rm g}$ is the mass-weighted average velocity of the gas particles relative to the BH particle. If instead, $v_{\theta}> v_{\rm bulk}$, we use 
\begin{equation}
    \dot{M}_{\rm Bondi}=4\pi\frac{(GM_{\rm BH})^2\rho_{\rm g}c_{\rm s}}{(c_{\rm s}^2+v_{\theta}^2)^{2}}
    \label{eq:bondi_3}
\end{equation}

We further limit the accretion rate to 10 times the Eddington accretion rate, defined as
\begin{equation}
\dot{M}_{\rm Edd} = 2.7\times10^{-3}\left(\frac{M_{\rm BH}}{10^5~\text{M}_\odot}\right)\left(\frac{\epsilon_0}{0.1}\right)^{-1}\rm~M_\odot~\text{yr}^{-1}\mbox{\ ,}
    \label{edd}
\end{equation}
where the radiative efficiency $\epsilon_0=0.125$ \citep{Negri2017}. $M_{\rm BH}$ is updated every timestep $\delta t$ as $\delta M_{\rm BH}=\dot{M}_{\rm Bondi}\delta t$. $M_{\rm dyn}$ tries to match $M_{\rm BH}$ growth by stochastically swallowing nearby gas particles in the BH kernel. 

We adopt a probabilistic procedure for BH sink particle swallowing of gas particles \citep{Springel2005_2}. The probability that a gas particle within the BH sink particle kernel will be swallowed (accreted) is 
\begin{equation}
    p_{\rm acc} = \frac{w\dot{M}_{BH}\delta t}{\rho_{\rm g}}\mbox{\ ,}
\end{equation}
where $w$ is the kernel weight of the gas particle and $\dot{M}_{\rm BH}$ the BH accretion rate. A random number $p$ is drawn, and if $p<p_{\rm acc}$ the gas particle will be swallowed.

\subsection{Black hole feedback}\label{sec:bhfeed}

We model thermal feedback from BHs \citep{Liu2020,Jeon2023,Jeon2024}, injecting energy into the gas particles in the BH kernel. The distributed energy for $\delta t$ is $\delta E = \epsilon_r L_{\rm BH}\delta t$, with the radiation-thermal coupling efficiency $\epsilon_r=0.02$ \citep{Tremmel2017}. $L_{\rm BH}$ is the BH luminosity defined as
\begin{equation}\label{eq:bhlum}
    L_{\rm BH} = \epsilon_{\rm EM}\dot{M}_{\rm BH}c^2\mbox{\ .}
\end{equation}
with c being the speed of light. The radiative efficiency $\epsilon_{\rm EM}$ is defined as
\begin{equation}
    \epsilon_{\rm EM} = \frac{\epsilon_0A\eta}{1+A\eta},~\eta\equiv\dot{M}_{\rm BH}/\dot{M}_{\rm Edd}\mbox{\ ,}
\end{equation}
with $A=100$ \citep{Negri2017}, asymptotically approaching $\epsilon_0$ for high accretion rates.

\begin{figure*}[!htb]
\gridline{
\fig{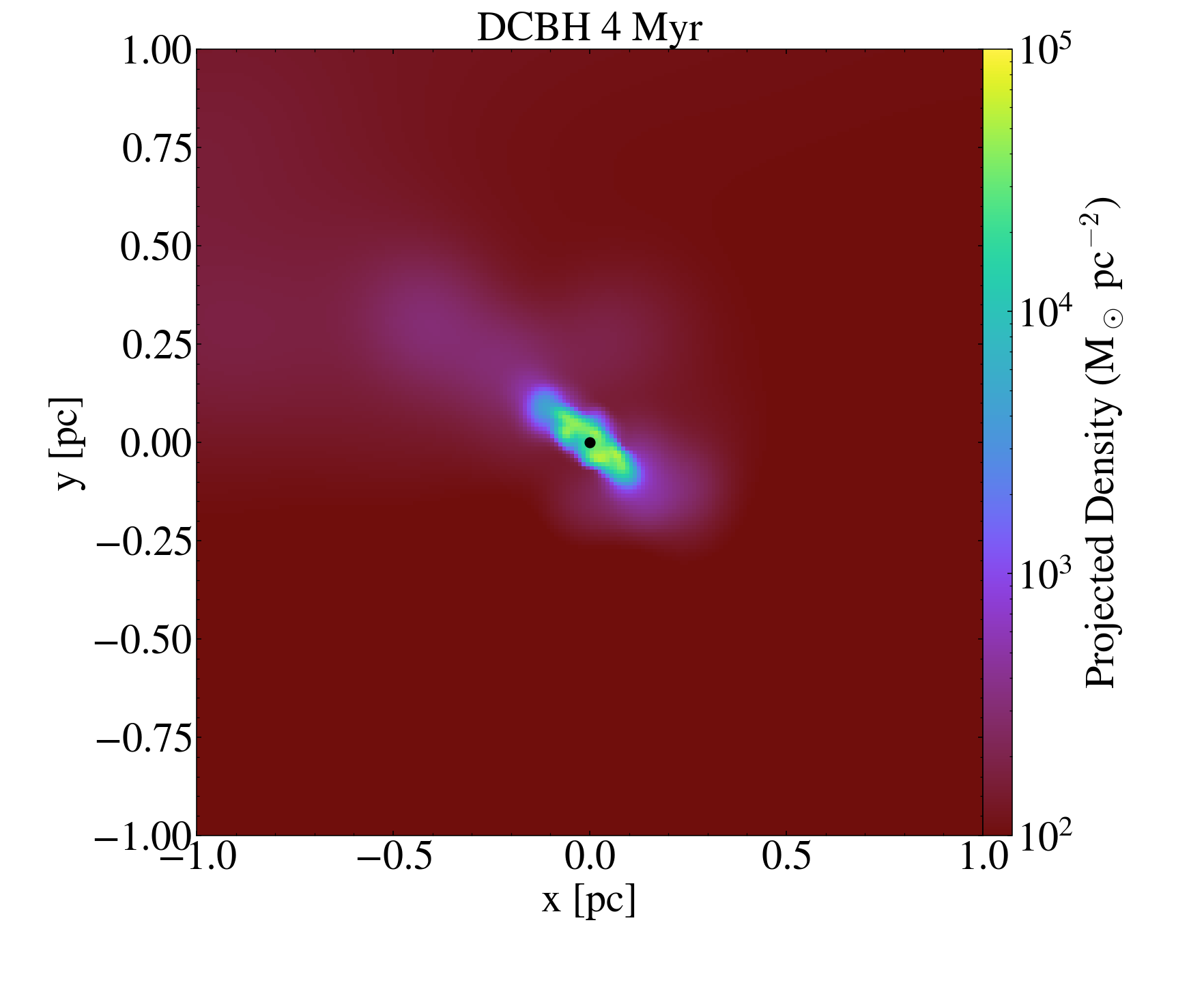}{0.51\textwidth}{}
\fig{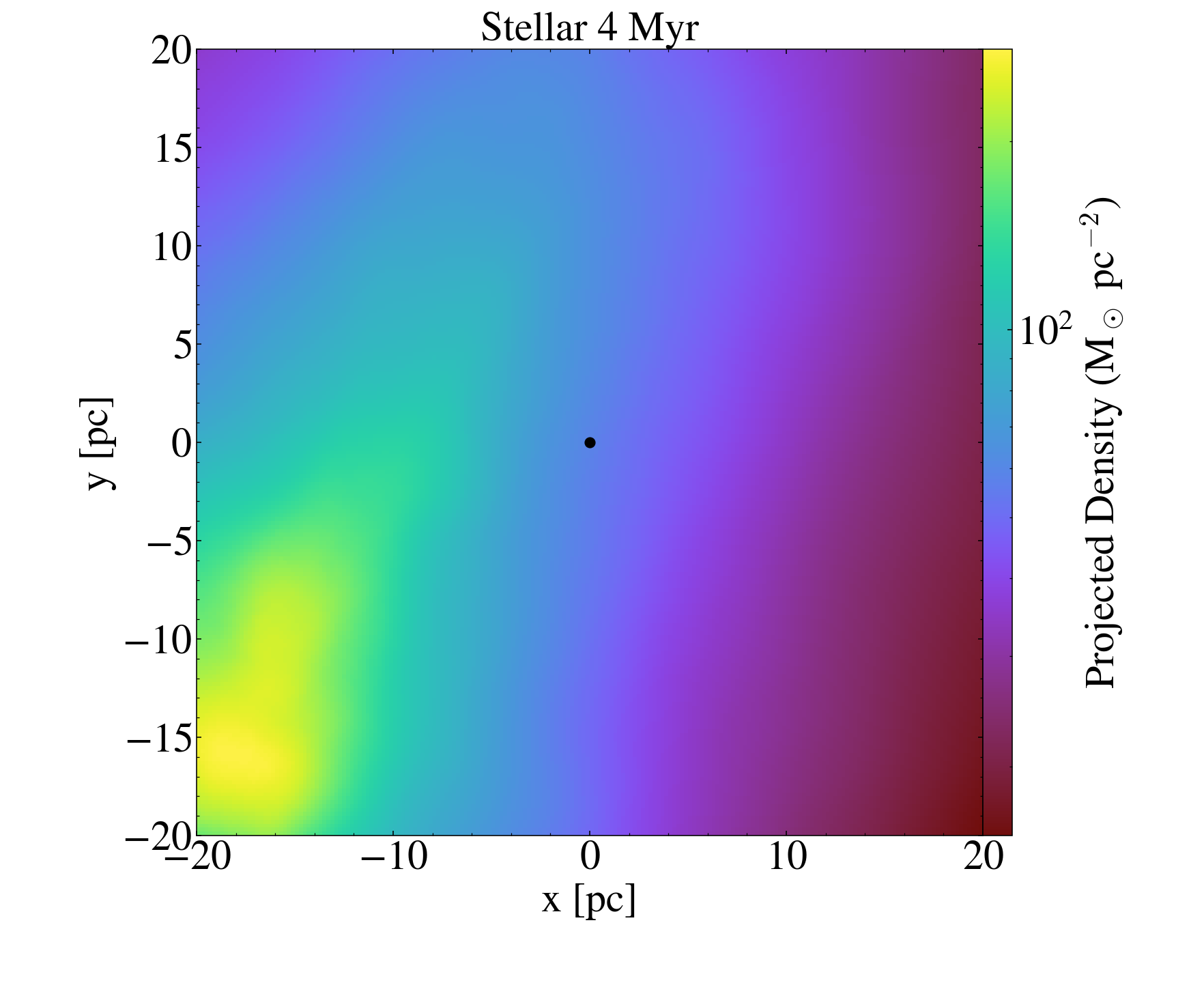}{0.5\textwidth}{}
}
\vspace{-4em}
\caption{Gas density projections near the seed BH in the cutouts after $\sim4$ Myr for the DCBH and Stellar runs. While the DCBH can amass high density gas in its vicinity because of its deep gravitational potential well, reaching up to $\sim3$ orders of magnitude above the ambient density, the much lighter stellar remnant BH cannot. As is also evident in Fig.~\ref{fig:ndist}, the heavy seed can cause dense gas to pile up near the BH, whereas the light seed is failing to do so, due to its shallower potential well and dynamical wandering.}
    \label{fig:projection}
\end{figure*}

\section{Results} \label{sec:results}

After running the zoom-in and cutout simulations, we post-process the simulation output with the \textsc{yt} \citep{Turk2011} and Meshoid \citep{Grudic_meshoid_2021} packages. We compare the gas environment at high resolution for BHs with different origins, see how BH accretion and feedback behaves close to the BH, and predict their spectra from the accretion and nearby gas environment.

\subsection{Gas distribution near the Black Hole}\label{sec:gasdis}

\begin{figure*}[!htb]
\gridline{
\fig{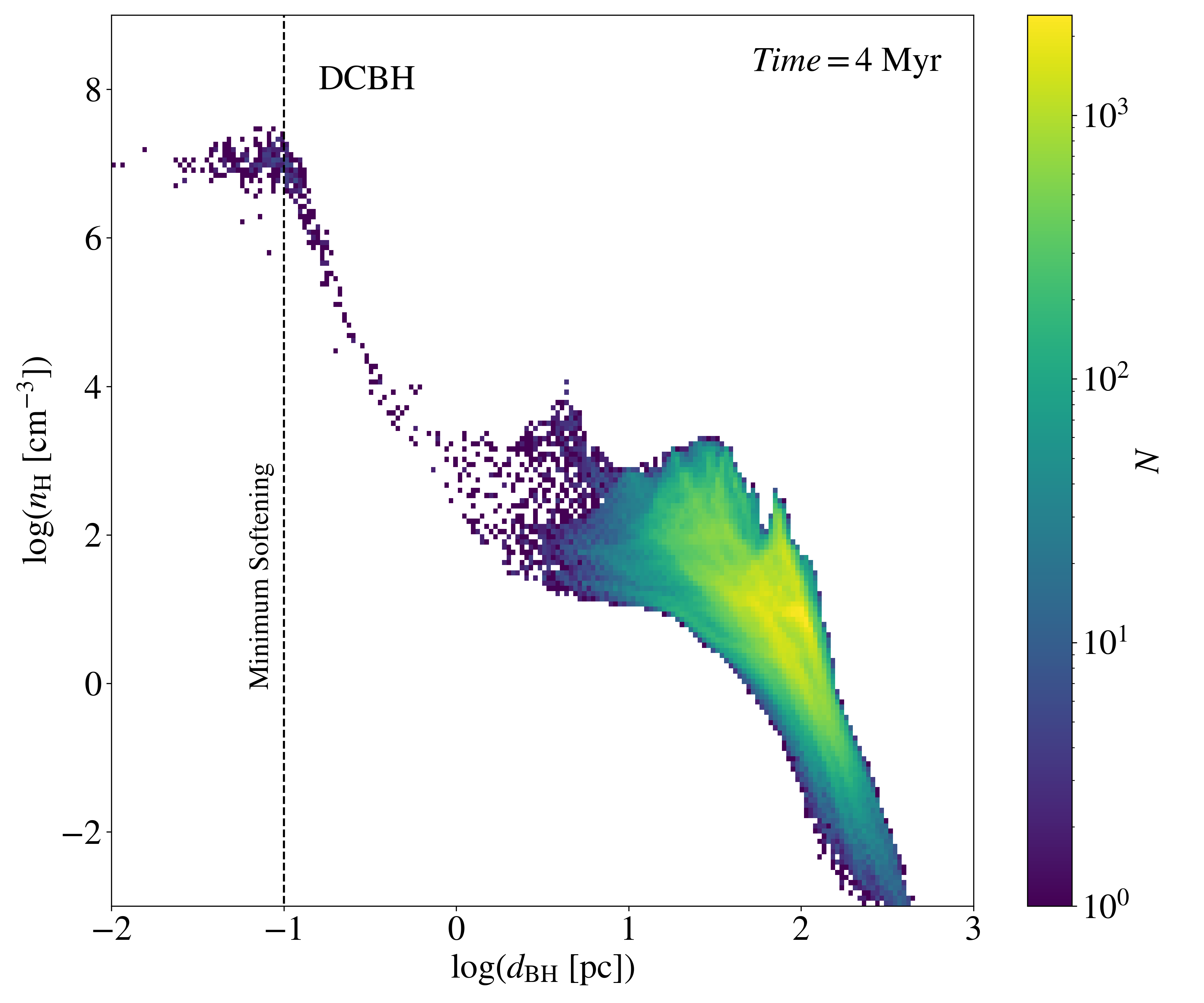}{0.5\textwidth}{}
\fig{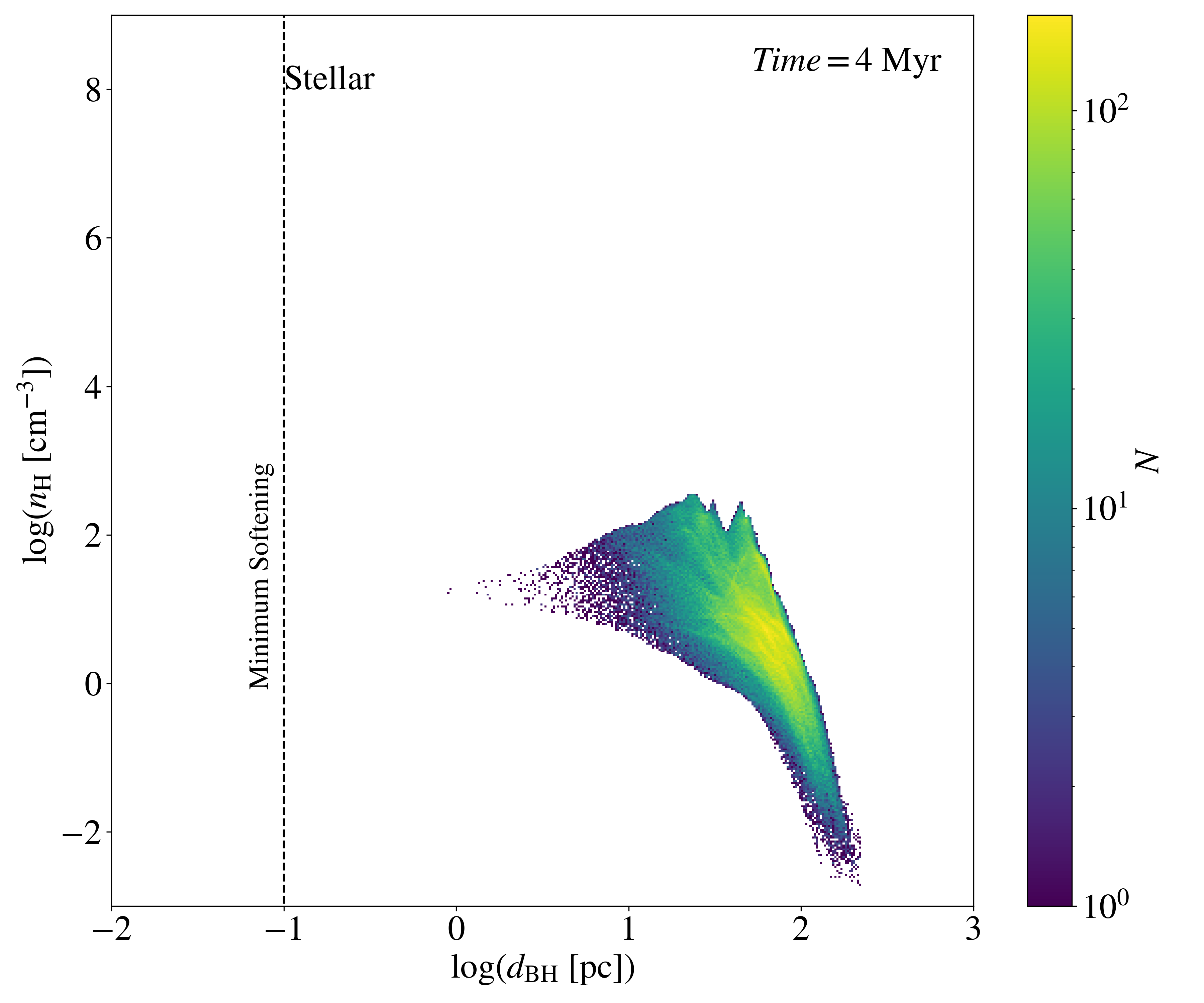}{0.5\textwidth}{}
}
\vspace{-3em}
\caption{Phase plots of hydrogen gas density versus distance from the BH in the cutouts after $\sim4$ Myr for the DCBH and Stellar runs. The massive DCBH affects the nearby gas more significantly, causing the gas to pile up at high densities in the vicinity of the BH, while the stellar remnant BH leaves the surrounding gas largely unaffected due to its low mass. The high-density gaseous cocoon created around the DCBH $(\sim10^7$ cm$^{-3}$), could produce LRD-like spectra \citep{Inayoshi2025_bl,Jeon2025_lrd,Naidu_BHstar2026}. 
}
    \label{fig:ndist}
\end{figure*}

We first examine the gas density and temperature evolution around the BHs in the cutouts after a few Myr. The gas density distribution is shown in Fig.~\ref{fig:projection}, where the DCBH resides within a high-density region, while the stellar remnant BH is located in a low-density, ambient gas environment. To further understand this behavior, in Fig.~\ref{fig:ndist} we display the hydrogen gas density distribution versus the distance from the BH sink particle after the cutouts have been run for $\sim4$\,Myr. We find that the massive DCBH can affect the gas in the cutout, to reach densities up to $\sim10^7$ cm$^{-3}$ when approaching the resolution limit at $\sim0.1$\,pc. In contrast, the light stellar seed cannot influence the gas significantly, so that the bulk of the gas is not influenced by the gravitational potential of the low-mass BH with no buildup of high-density gas as in the DCBH case.

\begin{figure*}[!htb]
\gridline{
\fig{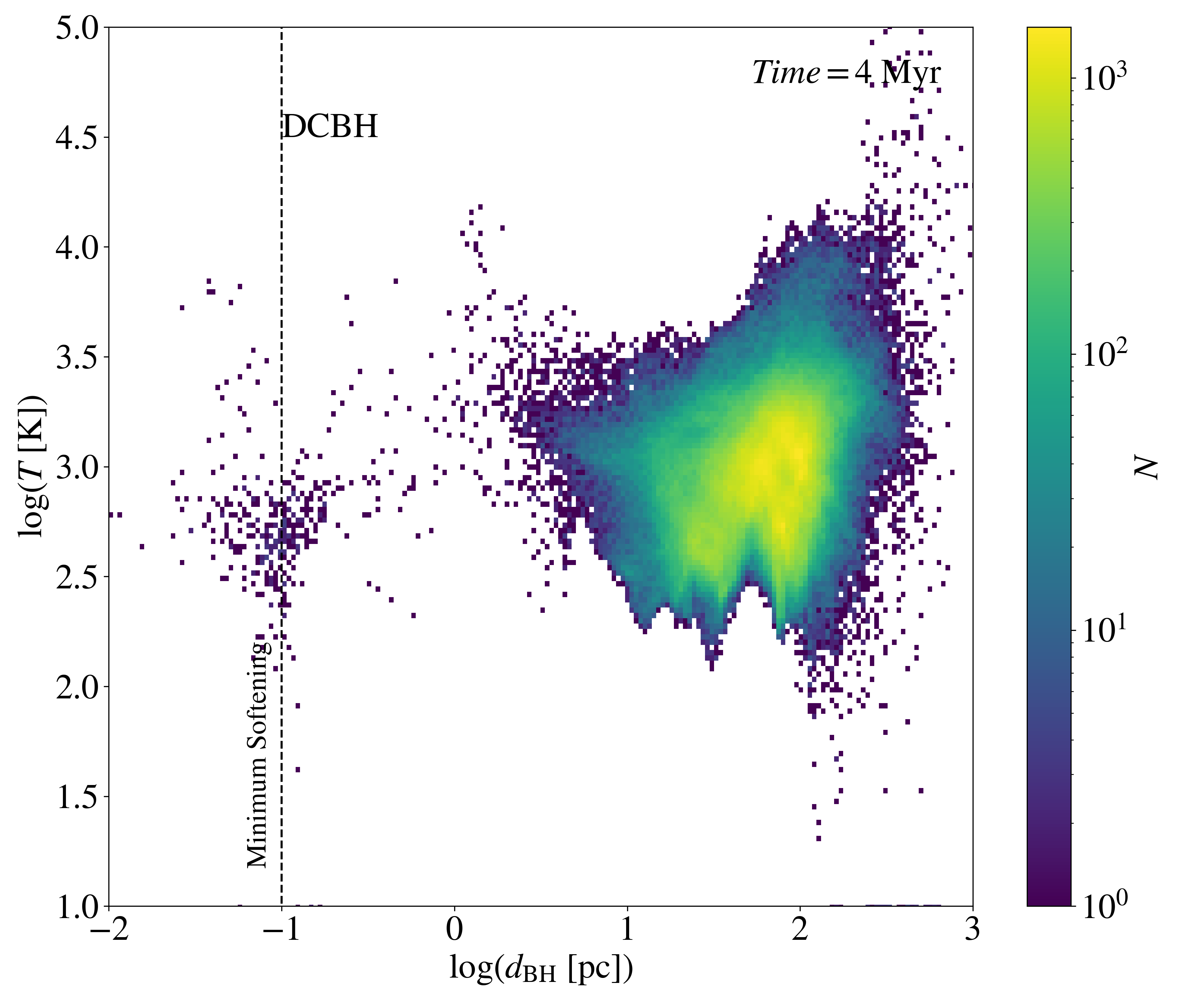}{0.5\textwidth}{}
\fig{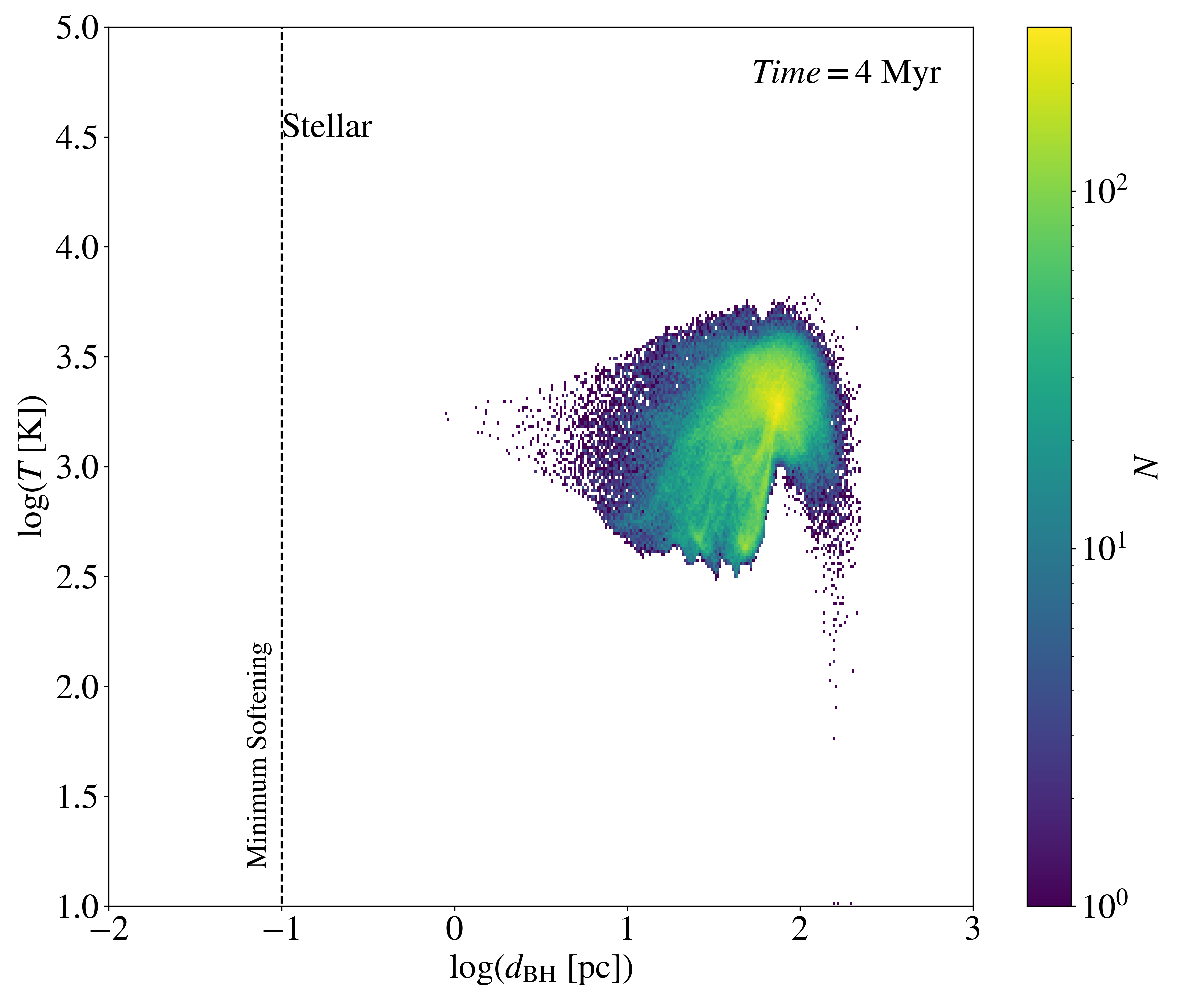}{0.5\textwidth}{}
}
\vspace{-3em}
\caption{Phase plot of hydrogen gas temperature versus distance from the BH in the cutouts after $\sim4$ Myr for the DCBH and Stellar runs. We do not find significant differences in the average temperature or its overall range between the two cases. We note that the DCBH cutout initially contained gas above $>10^4$\,K, heated by SN feedback, but such high temperature gas did not prevent the dense gas pileup of $\sim10^{2.5}$\,K gas near the DCBH.}
    \label{fig:Tdist}
\end{figure*}

Comparing the resulting gas temperature distribution in Fig.~\ref{fig:Tdist}, we find no significant differences in the temperature range 
between the DCBH and stellar remnant cutouts, both exhibiting average temperatures of $\sim10^3$\,K. We note that for the DCBH case, in difference from the substantially higher densities reached, the temperatures of the gas coalescing close to the BH are similar to those of the more widely distributed gas. In fact, the DCBH cutout region contains higher temperature gas ($>10^4$\,K) from SNe in the original Zoom simulation. Still, despite the presence of such hotter gas, the DCBH is able to trigger a dense gas buildup. 

\begin{figure*}[!htb]
\gridline{
\fig{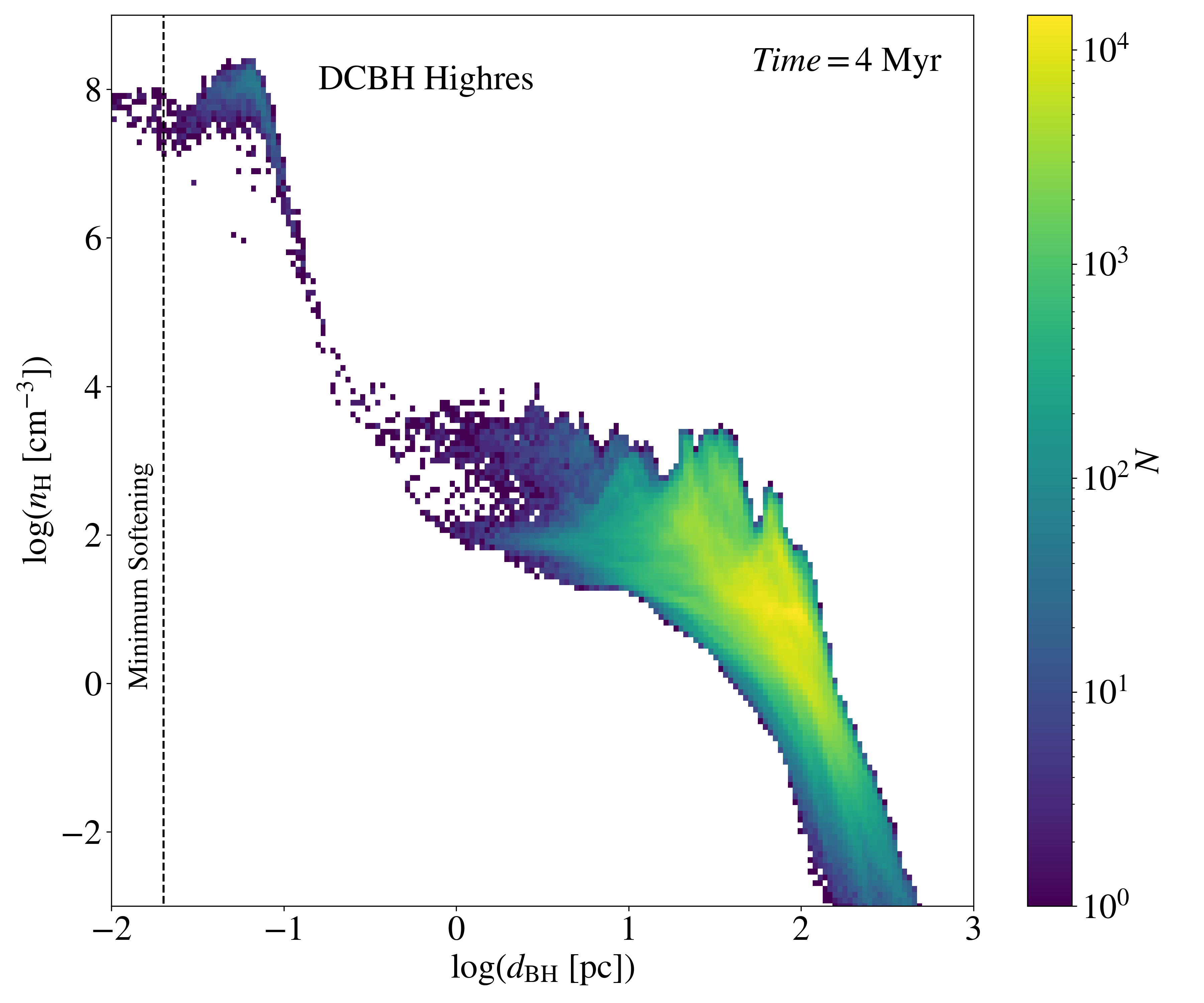}{0.5\textwidth}{}
\fig{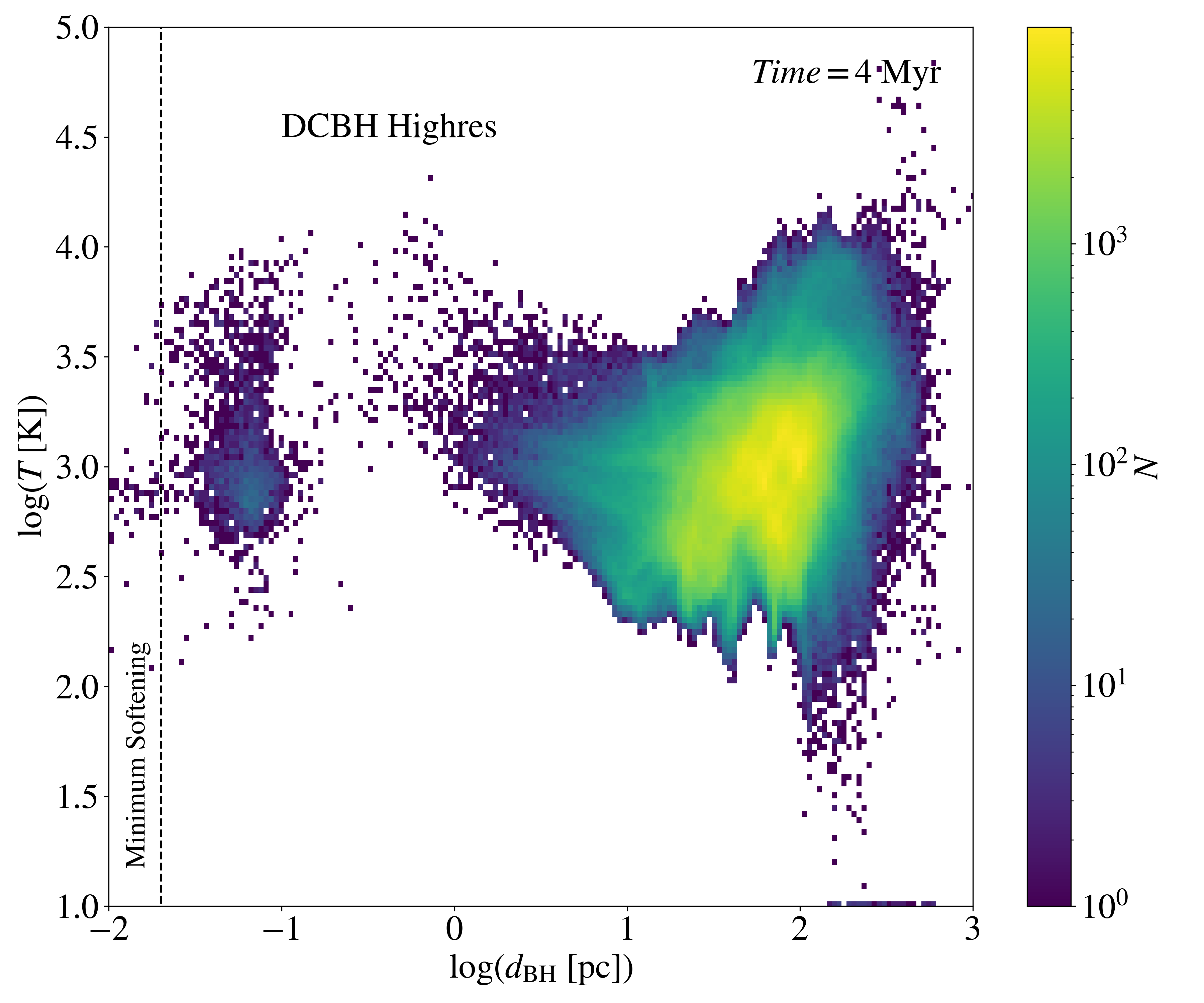}{0.5\textwidth}{}
}
\vspace{-3em}
\caption{Phase plots of hydrogen gas density and temperature versus distance from the BH for the DCBH\_Highres run after $\sim4$ \,Myr. Compared to the fiducial DCBH cutout run (left panels of Fig.~\ref{fig:ndist} and \ref{fig:Tdist}), the 5 times higher resolution allows the gas to reach even higher densities near the BH, up to $\sim10^8$ cm$^{-3}$, while the temperature distribution does not change significantly, other than comprising a larger number of gas particles closer in.}
    \label{fig:disthighres}
\end{figure*}

We note that the DCBH and Stellar cutouts are two distinct environments, drawn from the Zoom simulation at different redshifts and locations. Thus, some environmental dependence may exist in the gas temperature and density distribution. However, the two cutout systems are similar, having comparable masses with only a factor of $\sim3$ difference (see Table~\ref{tab:cutouts}), and with similar values of the virial parameter ($\alpha$), implying that the overall thermal energy content of the two systems is also similar. Thus, the environmental differences should not be dominant in determining how the gas responds to the central BH. 

\begin{figure*}[!htb]
\gridline{
\fig{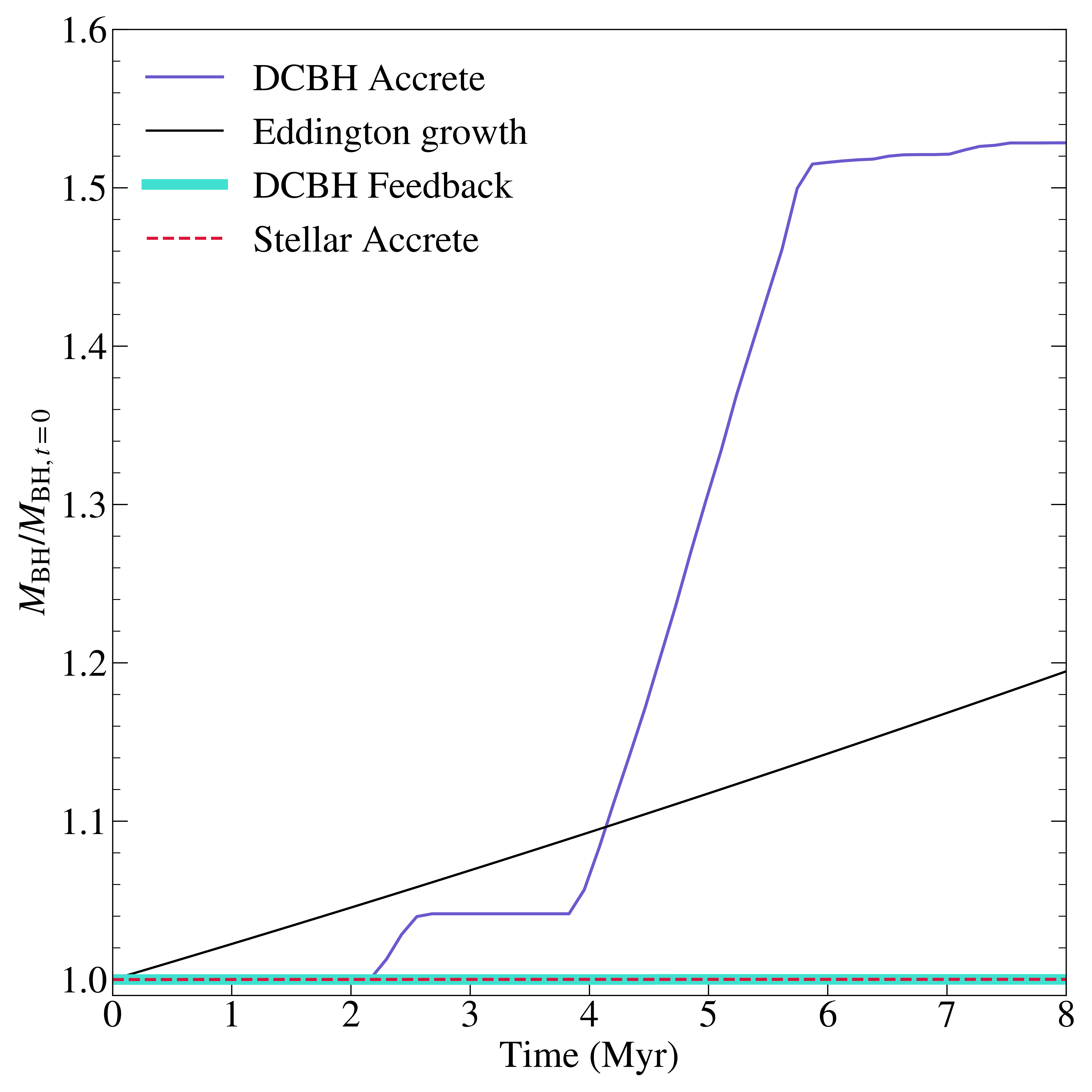}{0.5\textwidth}{}
\fig{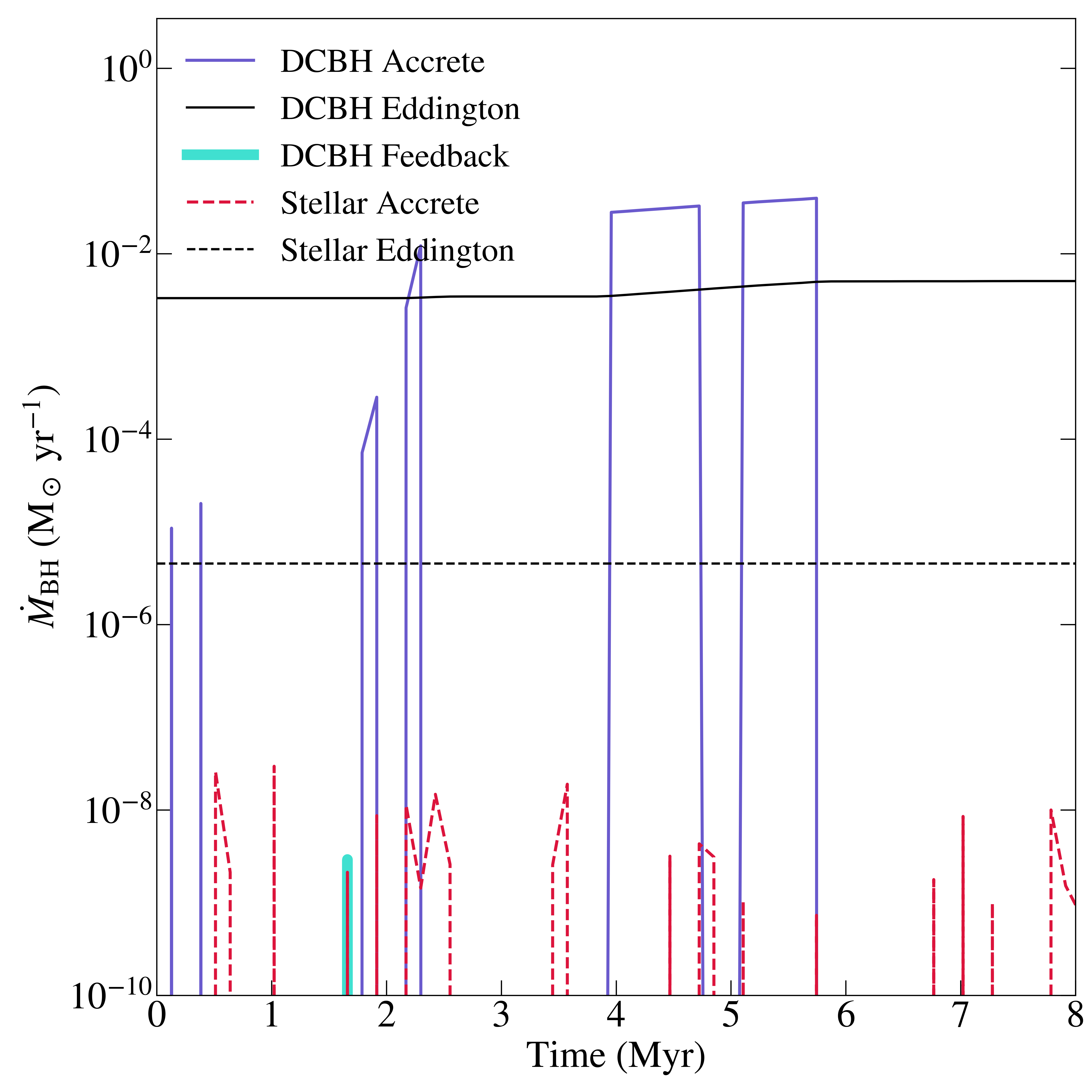}{0.5\textwidth}{}
}
\vspace{-3em}
\caption{BH mass growth and accretion rate for DCBH\_Accrete, DCBH\_Feedback, and Stellar\_Accrete cutouts. The left panel shows the ratio of current to initial BH mass, while the right panel displays the corresponding accretion rates. For comparison, we also indicate the BH mass growth under Eddington accretion, and the Eddington rates for the DCBH and stellar BH.
Only the DCBH\_Accrete case shows significant BH mass growth over $\sim8$ Myr, which occurs over several super-Eddington burst periods, as seen in other high-resolution simulations \citep[e.g.,][]{Mehta2026,Prole2026}. For light BH seeds, or with BH feedback enabled, accretion rates never reach the Eddington value, and BH growth is suppressed. However, the effects of BH feedback may be overestimated, due to its numerical implementation in the cutout region (see the main text for further discussion). 
}
    \label{fig:bhacc}
\end{figure*}

To assess whether numerical resolution has any effect on the results, we examine the gas distribution in the DCBH\_Highres cutout. Fig.~\ref{fig:disthighres} shows the gas density and temperature distributions versus the distance to the BH in the DCBH\_Highres run. The gas behaves similarly to the lower resolution case, except that the gas density near the BH reaches even higher values, up to $\sim10^8$ cm$^{-3}$. The higher resolution is able to resolve the buildup of dense gas even closer to the central BH seed, but overall, the difference in resolution does not significantly affect the resulting gas distribution. 

\subsection{Black hole accretion and feedback}\label{sec:bhaccandfeed}

The cutouts discussed above disabled BH accretion and feedback to first examine how increased resolution affected gas particle interaction with the BH. Now, we enable BH accretion and/or feedback to see how the higher resolution affects BH evolution. Fig.~\ref{fig:bhacc} shows the BH mass evolution and mass accretion rates for the DCBH\_Accrete, Stellar\_Accrete, and DCBH\_Feedback cutouts.


As expected from the gas distribution near the BHs (Sec.~\ref{sec:gasdis}), the DCBH\_Accrete cutout grows significantly in BH mass, able to swallow the high density gas nearby in brief periods of super-Eddington accretion. Such bursty super-Eddington accretion has previously been seen in other high-resolution simulations of SMBH growth \citep[e.g.,][]{Mehta2026,Prole2026,Chon2026}. Moreover, large-scale simulations that cannot fully resolve SMBH accretion have introduced artificial burstiness to match observational constraints on AGN luminosity functions \citep{Shen2026}. In contrast, the stellar remnant experiences sub-Eddington accretion rates throughout, resulting in insignificant mass growth. Therefore, we find that only heavy BH seeds could facilitate such bursty growth, whereas the light stellar remnant BHs cannot draw in the nearby gas, thus never experiencing high accretion episodes during the cutout run period, with maximal Eddington ratios of $\dot{M}_{\rm BH}/\dot{M}_{\rm Edd}\sim10^{-2}$ \citep[see also][]{Jeon2023}.

When we include BH feedback, the growth is suppressed even in the DCBH case. After a brief sub-Eddington accretion episode at $\sim2$ Myr, the nearby gas is evacuated from the vicinity of the BH and growth stops. 
However, the effects of BH feedback are likely overestimated in the cutout setting, as we do not properly model the radiation-hydrodynamical evolution of the initially compact `(thermal) feedback bubble' into the highly-anisotropic surrounding medium, but simply inject feedback energies to the BH kernel (Section~\ref{sec:bhfeed}). In reality, photon trapping and advection may act to blunt the impact of feedback (see Sec.~\ref{sec:caveat} for a discussion of numerical limitations).  
In contrast, previous studies have suggested that in thick disks enabling super-Eddington accretion, the effects of feedback may be mitigated \citep{Inayoshi2016,Jiang2019,Park2020}. Thus, our DCBH\_Accrete cutout might be more representative of the cases where super-Eddington accretion is possible with less efficient feedback. 

\begin{figure*}[!htb]
    \centering
    \includegraphics[width=0.9\textwidth]{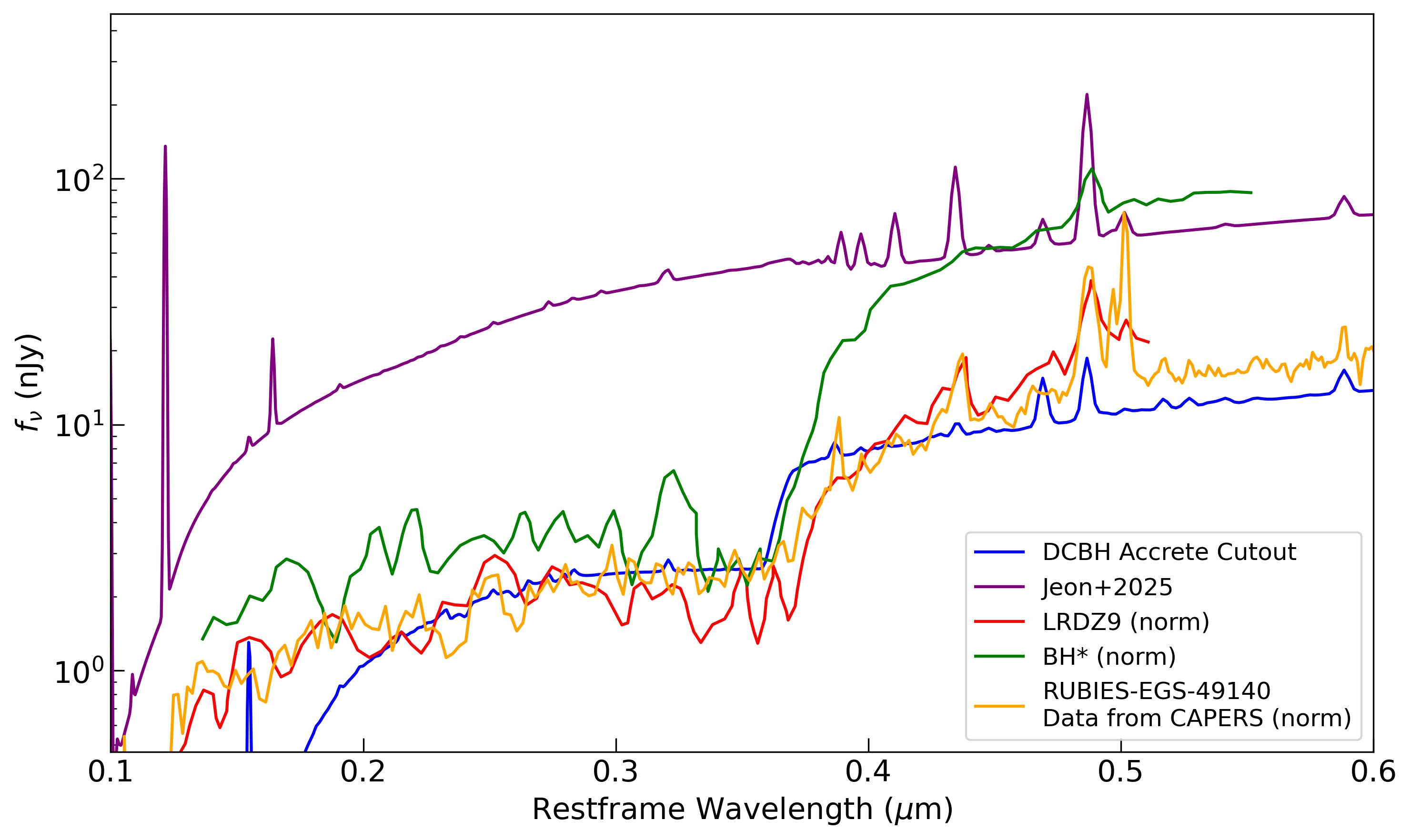}
    \vspace{-1em}
    \caption{Model spectra for the case of a strongly-accreting DCBH. We use results from the DCBH\_Accrete cutout run, taken at $\sim5.7$\,Myr, during a super-Eddington accretion episode as input for \textsc{Cloudy}. The spectral calculation uses the BH accretion rate and close-in gas structure from the cutout, resolved to $\sim0.1$\,pc. We compare our prediction against normalized observed LRD spectra \citep{Dickinson2024,Degraaff_2025_rubies,Taylor2025,Naidu_BHstar2026}, as well as a model DCBH spectrum from previous work, where the gas distribution near the BH was not properly represented, given a resolution of $\sim300$\,pc \citep[][]{Jeon2024}. With the cutout technique, on the other hand, the high-density region around the BH can be more fully resolved, which produces the strong Balmer breaks seen in observed LRD spectra.}
    \label{fig:spectra}
\end{figure*} 

\subsection{AGN Spectra}

To infer how the DCBH\_Accrete cutout will appear observationally, we use the \textsc{Cloudy} photoionization code, version C23.01 \citep{Chatzikos2023}. In previous work, we did not have information on the close-in gas distribution around a SMBH, given the limited resolution in the larger-scale cosmological simulation considered there. We, therefore, had to make simplifying assumptions in extrapolating to the unresolved scales, such as constant gas density or pressure \citep{Jeon2024}. In the cutouts, on the other hand, the nearby gas distribution is well resolved to sub-pc scales. 
We specifically derive \textsc{Cloudy} spectra for the gas density profile over the $\sim0.1$ to $\sim100$\,pc range of distances from the central SMBH (averaging at each radial bin), as found in the DCBH\_Accrete cutout run at $\sim5.7$ \~Myr, when the highest accretion rate is encountered. We thus avoid arbitrarily assuming a constant gas density for the previously unresolved scales.

For the AGN part of the broad-band spectrum, we use the \textsc{Cloudy} model implementation of an AGN with a multicolor disk UV component and a non-thermal X-ray component\footnote{We here include the X-ray term for completeness, but we do not further consider the X-ray band in our analysis, given that most JWST AGN have been observed to be X-ray weak \citep{Akins2024,Kocevski2023,Liu2026}.}:

\begin{equation}\label{eq:spectra}
    f_{\nu}\propto
  \nu^{-0.5}\exp\left(\frac{-h\nu}{k_{\rm B}T_{\rm disk}}\right)\exp\left(\frac{-0.136~\rm{eV}}{h\nu}\right)
   +a\nu^{-1}
\end{equation}

\noindent where $h$ is the Planck constant, $k_{\rm B}$ the Boltzmann constant, and $T_{\rm disk}$ the SMBH accretion disk temperature. The $a$-parameter is chosen to fix the X-ray to UV flux ratio, where the non-thermal term $a\nu^{-1}$ is only included when $h\nu\geq1.36$ eV. For the disk temperature, we use the slim disk model of \citet[][see their Eq.~20]{Watarai2006} for super-Eddington accretion, determining $T_{\rm disk}\simeq10^6$\,K. For the X-ray to UV ratio, we use the model of \citet[][see their Fig.~8]{Avni1986} and determine a ratio of $\sim-1.4$ in log scale. We employ the SMBH accretion rate at 5.7 Myr and Eq.~\ref{eq:bhlum} to determine the AGN luminosity as $\sim10^{44.4}$\,erg\, s$^{-1}$. Lastly, we use the average gas phase metallicity of the gas particles in the cutout, assuming that no significant metal enrichment has occurred since the cutout was initialized, with a value of $\sim0.02$ Z$_\odot$. 

The resulting spectra are shown in Fig.~\ref{fig:spectra}. We further show in comparison the model spectrum from a zoom-in cosmological simulation with lower resolution \citep{Jeon2024}, as well as normalized observed LRD spectra, including the BH* object \citep{Naidu_BHstar2026}, CAPERS-LRD-z9 \citep{Taylor2025}, and RUBIES-EGS-49140, with spectral data from the CANDELS-Area Prism Epoch of Reionization Survey \citep[CAPERS;][]{Dickinson2024,Degraaff_2025_rubies}. Whereas both simulation spectra show a red AGN continuum, the strong Balmer-break feature, similar to that of observed LRDs, can only be seen when the high-density gas in the vicinity of the central BH is properly resolved. Here, the required high resolution is achieved through the cutout methodology. 

\section{Discussion}\label{sec:discussion}
 On the basis of the BH evolution resulting from the cutout simulations, we consider the implications for the overall SMBH mass evolution and the possible origin of the compact LRDs. 

\subsection{Supermassive Black Hole Formation}\label{sec:bhgrowth}

To form the observed SMBHs with masses $\sim10^8$ M$_\odot$ at $z\gtrsim6$, the seed BHs must grow rapidly, at near- or super-Eddington rates \citep[e.g.,][]{Smith2019_2,Bhowmick2024,Jeon2024,Mehta2026}. We find that such growth is only possible for heavy seeds. As shown in Fig.~\ref{fig:bhacc}, only the DCBH cutout experiences significant bursty growth and super-Eddington accretion periods. Without having an initially large mass and deep gravitational potential well, the light BH seed could not gather the nearby gas (Fig.~\ref{fig:ndist}), and the resulting growth was negligible. 

This conclusion would imply that most of the observed high-redshift SMBHs originated from heavy seeds. While other channels such as BH mergers exist for BH mass growth besides gas accretion, previous studies have found that accretion significantly contributes to SMBH growth even when considering high merger rates \citep{Bhowmick2025,Bhowmick2026_acc}. Furthermore, there are multiple pathways for heavy seed formation in addition to the DCBH scenario, which is our main focus here, such as hierarchical mergers (see Sections~\ref{footnote} and \ref{sec:firecomparison}).
If the various heavy seed pathways place the BH into a largely similar environment, allowing for a period of time to lose memory of the specific seeding process, our cutout simulations would then be expected to exhibit similar behavior for the subsequent evolution, provided that the main factor determining the gas distribution and the resulting BH accretion is the BH mass, not the seed formation mechanism. 
However, the DCBH pathway naturally places a massive seed in a pristine and cold gas environment lacking stars, and may be more effective in inducing dense gas buildup and super-Eddington accretion. Scenarios like hierarchical mergers, on the other hand, will form heavy seeds within dense stellar environments, such that the gas may be heated and/or expelled by stellar feedback even in the presence of a deeper BH gravitational potential well.

To derive a rough estimate for the heavy BH seed mass needed to allow for efficient gas accretion, we consider the Bondi radius, giving the BH's sphere of gravitational influence:
\begin{equation}\label{eq:br}
    R_{\rm Bondi} = \frac{2GM_{\rm BH}}{c_{\rm s}^2} \mbox{\ .}
\end{equation}
To calculate $c_s$, we assume an average gas temperature of $10^3$\,K near the BH (Fig.~\ref{fig:Tdist}). We obtain the mass enclosed within the BH's region of gravitational influence by assuming a sphere of radius $R_{\rm Bondi}$, containing hydrogen gas with average number density $100$ cm$^{-3}$ (Fig.~\ref{fig:ndist}). 
We find that for a $10^5$ M$_\odot$ BH, the mass within the Bondi radius is $\sim5\times10^6$ M$_\odot$, while for a $100$ M$_\odot$ BH the mass enclosed is only $\sim5\times10^{-3}$ M$_\odot$. For the intermediate case of a $10^4$ M$_\odot$ BH, the enclosed mass is $\sim5\times10^3$ M$_\odot$. We recognize that the enclosed mass estimate for the $10^5$ M$_\odot$ BH is larger than the mass contained in the cutouts (Table~\ref{tab:cutouts}), so that the single gas density and temperature approximations will break down at larger scales. Still, this argument implies that to enable efficient gas accretion onto a BH, heavy BH seeds of $\sim10^5$ M$_\odot$ are initially needed to influence a sufficiently large gas mass, comparable to or larger than its current mass, and build up a high-density region as shown in Fig.~\ref{fig:ndist}. As long as the gas reservoir can be replenished from the cosmic environment, the heavy seed will be able to accrete and grow significantly in mass. In a previous study, we  suggested that such replenishment is possible for the biased halos, where DCBHs originate \citep{Jeon2024}. 

As massive quasars and high-redshift AGN are relatively rare, it would not be surprising that they originate from heavy BH seeds that also require rare conditions for formation. To fully characterize the number density of heavy BH seeds, large volume simulations or semi-analytic models are needed \citep[e.g.,][]{Habouzit2016,Jeon2025,Porras2025,Bhowmick2024}, enabling the comparison with the observed SMBH population statistics \citep{Taylor2024,Fei2025}. Previous work has indicated that rare and biased halos are able to support efficient assembly of SMBHs at high-redshifts \citep{Dimatteo2012,Dubois2012,Kiyuna2023,Jeon2026_ff,Kho2026,Kiyuna2025}, which implies that clustering measurements of SMBH could provide further clues to their origin.

\subsection{Little Red Dots}

Regarding LRDs, systems proposed to be AGN with subdominant stellar populations, the spectra derived from the cutout runs, only harboring an AGN component, could explain the red part of the LRD spectrum, as shown in Fig.~\ref{fig:spectra}. The blue part cannot be fully explained without a stellar population, which we have left out in our cutout runs. 

The dense gas region triggered by the massive BH produces the Balmer breaks in the spectrum at $\sim0.35$ $\micron$ comparable to the observed LRDs. However, for the most extreme LRDs like the BH* object \citep{Naidu_BHstar2026}, their extreme Balmer breaks cannot be reproduced. Even higher gas densities than those achieved in the cutouts may be needed. Higher resolution or a more massive SMBH could accomplish this, as higher resolution in the cutout, up to 0.02 pc, led to higher gas densities near the BH, up to $10^8$ cm$^{-3}$ (Fig.~\ref{fig:disthighres}). In addition, the dense gas regions, proposed to be the source of the strong Balmer break in LRDs, are thought to have sizes of $\sim1000$ AU \citep{Inayoshi2025_lrd,Kokorev2025}, corresponding to $5\times10^{-3}$ pc. Thus, even in our cutouts, we have not yet reached the scale of the dense-gas regions in LRDs (see Fig.~\ref{fig:enclosed}). In future work with even higher resolution, we may be able to directly capture the expected scale of the dense LRD gaseous cocoon, thus reproducing the most extreme cases.

We note that because of stochastic BH accretion, corresponding to intermittent periods when the BH is luminous with high $\dot{M}_{\rm BH}$, the LRD phase may also be variable \citep{Lambrides2026}. However, if the SMBH is enshrouded in a dense gas envelope as some LRD models suggest, full radiative transfer calculations will have to be performed to measure how long the gas envelope remains bright, even after BH accretion ends, and to determine its variable nature.  

In any case, we find that similar to SMBH growth, heavy BH seeds are necessary to induce the dense gas envelope, and to produce LRD-like spectra for a central AGN engine. Other works have suggested that SMSs or early DCBH phases soon after formation could be the origin of LRDs, as they naturally produce a system with a gas envelope surrounding an accreting compact central object \citep{Chisholm2026,Kokorev2026_sms,Whalen2026,Pacucci2026}. We complement such studies to show that to sustain an LRD-like phase after BH seed formation, a heavy BH seed is required. Light BH seeds, on the other hand, corresponding to stellar remnants, will be unable to grow in mass, as indicated by our cutout simulations, and thus cannot form a dense gas region or LRD-like systems. 

\subsection{Comparison with Previous works}\label{sec:firecomparison}

We compare our results to select studies with a similar focus on early BH growth in a cosmological context, without aiming for completeness, to gauge the results' robustness in terms of numerical methodology and different assumptions for the sub-grid physics.  

Given that they also use the \textsc{gizmo} code, it is instructive to consider
the FORGE'd in FIRE simulations \citep{Hopkins2023,Hopkins2024,Hopkins2024_2}. This simulation suite resolves the SMBH accretion disk with even higher (AU-scale) resolution within a larger cosmological simulation, albeit at high computational cost. We generally find consistent results with FORGE'd in FIRE. They resolve a region around a $\sim10^7$ M$_\odot$ SMBH, and find a similar build-up of gas density and temperature \citep[$\sim10^8$ cm$^{-3}$ and $\sim10^3$ K at $\sim0.1$ pc;][]{Hopkins2023}. We note that FORGE'd in FIRE finds that SF occurs down to $\sim10^{-2}$~pc away from the BH \citep{Hopkins2023,Hopkins2024_2}. However, they further measure SF suppression within $10\sim100$ pc, inside the region where the BH gravitational potential dominates. 
With very low expected SF efficiencies, the possibly unresolved SF at the smallest scales should not significantly impact BH accretion.
In addition, FORGE'd in FIRE does not find a significant impact from magnetic fields at larger scales. At sub-pc scales, magnetic fields stabilize the accretion disk against fragmentation and enhance the gas inflow near the BH, but these effects are present  at $\lesssim10^{-1}-10^{-2}$ pc scales  \citep{Hopkins2023,Hopkins2024}, below our resolution limit. The absence of magnetic fields in our cutout simulations thus should not affect our conclusions for the dense gas buildup and BH mass evolution. 

The BRAHMA simulations \citep{Bhowmick2024_2} found that hierarchical mergers of lower mass ($\sim10^3$\,M$_\odot$) seeds could produce BHs of $\sim10^5$\,M$_\odot$, providing an alternative pathway for heavy seed formation. However, they find that the $10^5$\,M$_\odot$ BHs also grow mainly by mergers until $\sim10^6$ M$_\odot$, then through gas accretion to higher masses \citep{Bhowmick2024,Bhowmick2026_acc}. The subdominant growth through accretion may arise from unresolved dense gas near the massive seeds due to the lower resolution in cosmological simulations (Fig.~\ref{fig:enclosed}), and the Bondi-Hoyle formalism underestimating accretion rates. Still, they reach qualitatively similar conclusions - that a massive BH is initially required to enable efficient accretion and produce the observed high-redshift SMBHs and quasars.

We note that other recent studies have found contrary results. The SEEDZ simulations \citep{Mehta2026}, employing extremely high resolution (in a separate, smaller volume run than shown in Fig.~\ref{fig:enclosed}), reach $\sim0.1$\,pc at the smallest scales, similar to this work. They conclude that light stellar-remnant seeds with initial masses of $\sim100$\,M$_\odot$ could rapidly grow to $\sim10^3-10^4$\,M$_\odot$ in bursts of super-Eddington accretion, even without mergers. Below, we consider possible reasons for the conflicting results. 

SEEDZ does not include radiative (non-SN) feedback from stellar particles before they collapse into a light BH seed. This choice is based on previous work finding that the dense gas around Pop~III stars traps the ionizing radiation, so that stellar feedback would not impact the subsequent evolution of the larger-scale environment \citep{Regan2020,Jaura2022}. However, these studies considered relatively limited time evolution ($\sim20$\,kyr), or had to post-process the ionizing radiation due to high computational costs. Other studies that include radiation-hydrodynamical effects have found contrasting results, where the (initially D-type) ionization front surrounding massive Pop~III stars eventually breaks out (becoming R-type) to reach beyond the host halo, thus severely limiting the accretion onto the BH remnants \citep{Whalen2004,Alvarez2006,Abel2007,Jeon2014}. 
Even if the progenitor star is initially located inside a dense gas region, the R-type ionization front after break-out (at $\sim60$ kyr for a $\sim100$\,M$_\odot$ Pop~III star) enables a champagne flow that significantly drives down the density. 
This low-density environment also makes the subsequent SN feedback more disruptive \citep[e.g.,][]{Greif2007,Ritter2012}. In contrast to SEEDZ, our simulations do include photo-ionization feedback from stellar particles before BH formation, which will significantly hinder subsequent BH growth \citep{Johnson2007_2,Jeon2023}. Therefore, the rapid stellar remnant mass growth may arise from the optimistic assumption of not including stellar radiative feedback.

Another possibility is that the difference arises from the short timescales and limited number of cases simulated in our work (see Section~\ref{sec:caveat}). The cutouts can only be run for short periods ($<10$~Myr) due to the artificial vacuum boundary conditions. Given that the SEEDZ simulation finds bursts of super-Eddington accretion that are separated by $\sim50-100$~Myr, our cutouts could represent a dormant period, and we might not have caught the rare conditions enabling a high-accretion phase. The SEEDZ study further concludes that only a small fraction of stellar remnants experience rapid growth (of order a few percent), whereas we only simulate a small number of light-seed cases. 
Our choice of a light seed could be one of the common dormant cases. Further work is needed to more fully explore this statistics.

\subsection{Caveats}\label{sec:caveat}

A number of caveats needs to be considered in this work. The cutout simulations, by their nature, can only be run for short periods of time. The larger cosmic environment will eventually affect the cutout region, after which the cutouts will be physically unrealistic. Thus, our analysis have been limited to $\lesssim8$ Myr. While we have examined the BH evolution over short periods, this behavior may change over cosmological timescales. One example concerns the gas reservoir replenishment. Fig.~\ref{fig:bhacc} shows short periods of super-Eddington accretion leading to BH growth, but for such behavior to continue, the accreted gas will have to be continuously replaced. More substantial gas replenishment will be needed as the BH mass grows to maintain super-Eddington bursts. Therefore, the BH evolution predicted in this work might not be long-lasting.

Furthermore, we have only examined one case of DCBH and stellar remnant BH in this work. We have chosen the most optimistic cases for BH growth based on gas environment (Sec.~\ref{sec:cutset}), but a diversity of BH evolution pathways may exist, reflecting environmental factors beyond the BH formation mechanism. However, both the limited time coverage and number of cases probed, do not change our conclusions that to produce the observed high-redshift AGN and LRDs, heavy BH seeds may be favored. How the gas reacts to BH mass/gravitational potential should remain consistent over long periods, and if the light BH seeds are unable to affect the nearby gas initially to grow in mass, they will not be able to grow through accretion over cosmological timescales either. Moreover, previous works have found that heavy DCBH seeds usually form in massive and biased halos, whose gas environment support efficient accretion to be able to produce the observed high-redshift SMBHs and overmassive systems \citep{Chon2024,Jeon2024,Chon2026}, so the short-term efficient mass growth we see for heavy seeds in cutouts could be sustained for longer periods, producing the observed high-redshift SMBHs. 

In addition, we are not able to fully encapsulate BH physics even with the cutouts. Our sub-parsec resolution still does not reach the inner accretion disk scales, as shown in Fig.~\ref{fig:enclosed}, such that the approximate Bondi-Hoyle formalism has to be used. While artificial boost factors are not needed, unlike in cosmological simulations with much lower resolution that cannot resolve the dense gas buildup near the BH, the BH accretion modeled in this work is still an approximation. The actual BH growth at the event horizon may be lower or higher than the growth inferred in this work. General relativistic magnetohydrodynamics simulations that directly model the accretion disk have found that the Bondi-Hoyle model actually overestimates BH accretion \citep{Cho2025,Su2026}. Another key aspect of BH physics we cannot fully model is the resulting feedback. As stated above (Sec.~\ref{sec:bhaccandfeed}), we currently do not accurately model the expansion of the (initial) compact thermal feedback bubble into the anisotropic surrounding medium. The feedback model built for lower resolution cosmological simulations imparts BH feedback energy to all particles within the BH kernel, while the realistic interaction between the (now resolved) dense anisotropic medium and the expanding BH feedback bubble will be more complex. Processes such as photon trapping and propagation of the feedback front are not modeled. Thus, the gas response in the cutouts to BH feedback may be overestimated. Future work will be needed to reach even higher resolution, explore a diversity of BH seed environments, and resolve this issue of consistently modeling BH feedback. 


\section{Conclusions}\label{sec:conclude}

To address the origin of high-redshift SMBHs and LRDs, we have run targeted hydrodynamical simulations within cosmological initial conditions. Specifically, we have used the cutout technique, where we excise a small region of the gas environment and resimulate it at a higher resolution for short periods to connect the large-scale cosmological environment to small-scale BH evolution. We explore two BH evolution scenarios, a DCBH heavy seed, and a stellar remnant BH light seed. The massive heavy seed with its deeper gravitational potential well is able to influence the nearby gas and build up a dense gas region around it. Such dense gas around the BH enables intermittent super-Eddington accretion, and may produce LRD-like spectra. The light seed, on the other hand, given its shallow potential well, cannot significantly affect the nearby gas. Consequently, there is no dense gas buildup around the BH, resulting in weak accretion and mass growth during the cutout run. Light seeds are thus expected to stay dormant in general, if they only grow through gas accretion. 


Our results suggest that rare heavy BH seeds are necessary and most likely the progenitors of high-redshift SMBHs and LRDs, whether they originate from a DCBH or a different heavy BH seed scenario. Follow-up work will be needed to further test this proposal, as our current study is exploratory and only considers individual cases. Other simulations have found contrary results where light BH seeds could grow significantly through gas accretion \citep[up to $\sim10^4$ M$_\odot$;][]{Mehta2026}, although still not to the heavy DCBH seed regime $(\sim10^5$ M$_\odot$), and invoking idealized assumptions such as neglecting stellar radiative (non-SN) feedback. As different simulations adopt different formalisms and subgrid models for BH evolution, future investigations will be needed to determine what causes such differences, whether they arise from true physical effects or numerical variations. The number densities and formation times of heavy BH seeds are also not well constrained, due to the uncertainty in their formation criteria \citep{Sugimura2014,Inayoshi2015,Chon2016,Chon2026}. Large volume simulations testing the different heavy BH formation scenarios will be necessary to compare the number of heavy seeds to the observed high-redshift SMBHs and LRDs, thus assessing whether sufficient heavy seeds could be produced to match the observations. 

In the future, to further elucidate the emergence of high-redshift SMBHs, observations at higher redshifts or reaching lower BH masses will be needed \citep{Habouzit2016,Pacucci2017,Natarajan2017,Wise2019,Volonteri2021,Jeon2025}. Previous studies have shown that by the time they are observed, SMBHs have lost their seeding signature \citep{Taylor2024,Jeon2025}. To discriminate SMBH origins, gravitational lensing can enable observations of $\sim10^5$ M$_\odot$ BHs at the predicted heavy seeding scale for DCBHs \citep{Fei2025}, thus providing direct confirmation of the DCBH scenario in the near-future. Upcoming wide-field surveys with the Roman Space Telescope, predicted to discover AGN at $z\sim6-10$ \citep{Zhang2024}, will be able to perform population level studies of high-redshift SMBHs and constrain the various theories of BH formation. Farther ahead, multi-messenger searches with gravitational wave observatories will constrain SMBH populations in novel and complementary ways \citep[e.g.,][]{LISA_Review2023}. 
The LISA observatory will detect binary systems to measure SMBH populations and the contribution of mergers to their formation \citep{Robson2019}, a channel not explored in this work. 

For LRDs, searches are now being extended to lower redshifts \citep{Loiacono2026,Kapoor2026}. As LRD observations have been focused at high-redshift, constraining the evolution of their physical properties and population number densities at later times will be crucial \citep{Kocevski2023,Ma2026}. Identifying the LRD population at low redshifts will also provide strong tests for the heavy seed scenario for LRDs \citep{Jeon2025_lrd,Cenci2026,Pacucci2026}, which have been invoked in part to address their apparent high-redshift abundance.

Overall, the discoveries of AGN and LRDs at high-redshift enabled by JWST, and in close synergy with theoretical modeling, have invigorated questions on the formation of the first SMBHs. These remarkable beacons of the early Universe may well hold the key to unlocking the physics of structure formation at Cosmic Dawn. 

%

\begin{acknowledgments}
The authors acknowledge the Texas Advanced Computing Center (TACC) for providing HPC resources under allocation AST23026. VB acknowledges support from the Josey Centennial Professorship in Astronomy at UT Austin. Claude Sonnet~5 was used to streamline the Python code that produced parts of the figures. All data and images are from original work by the authors.
\end{acknowledgments}

\appendix
\section{Numerical test of the cutout method}\label{sec:appendix}

\begin{figure*}[!htb]
\gridline{
\fig{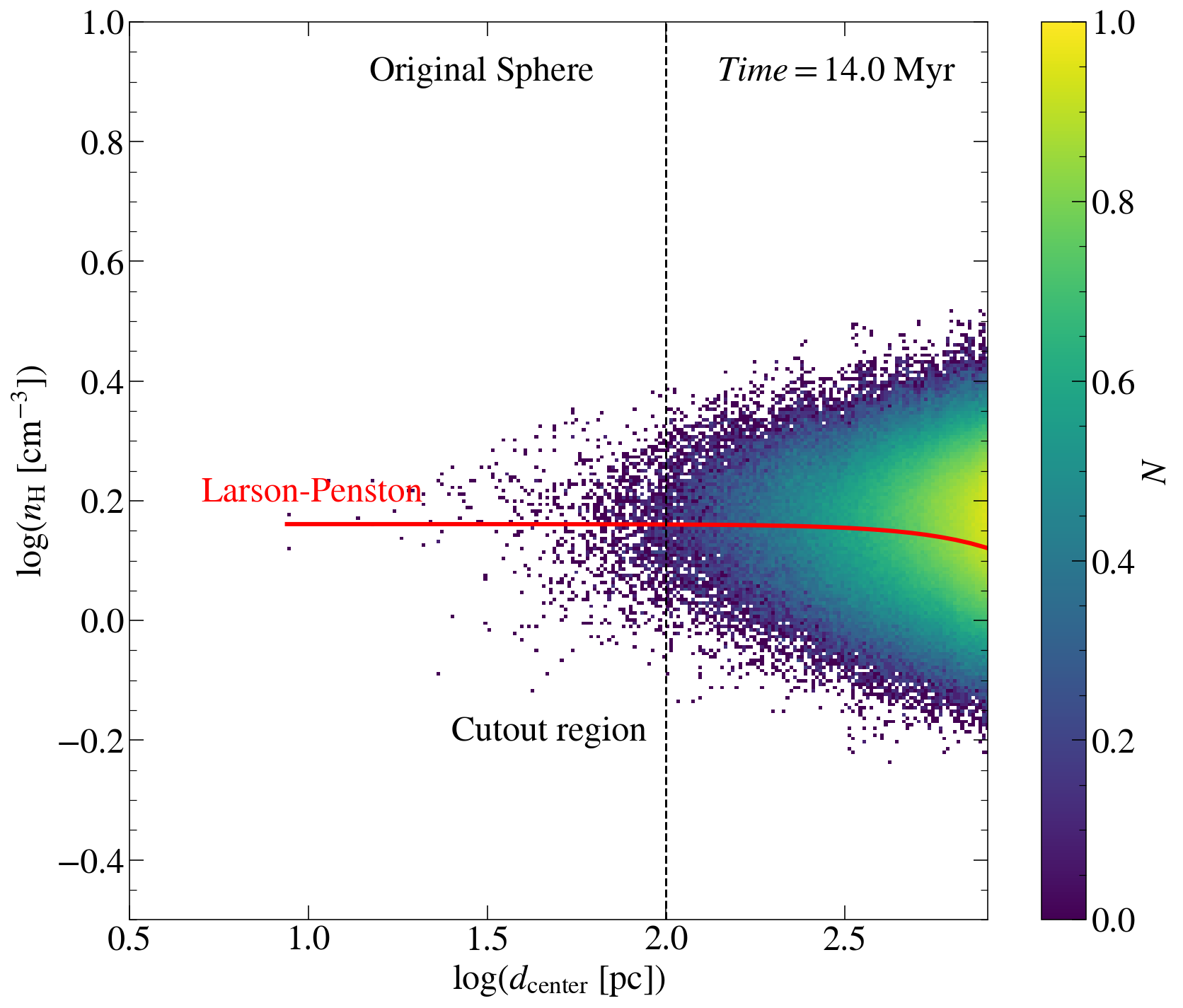}{0.5\textwidth}{}
\fig{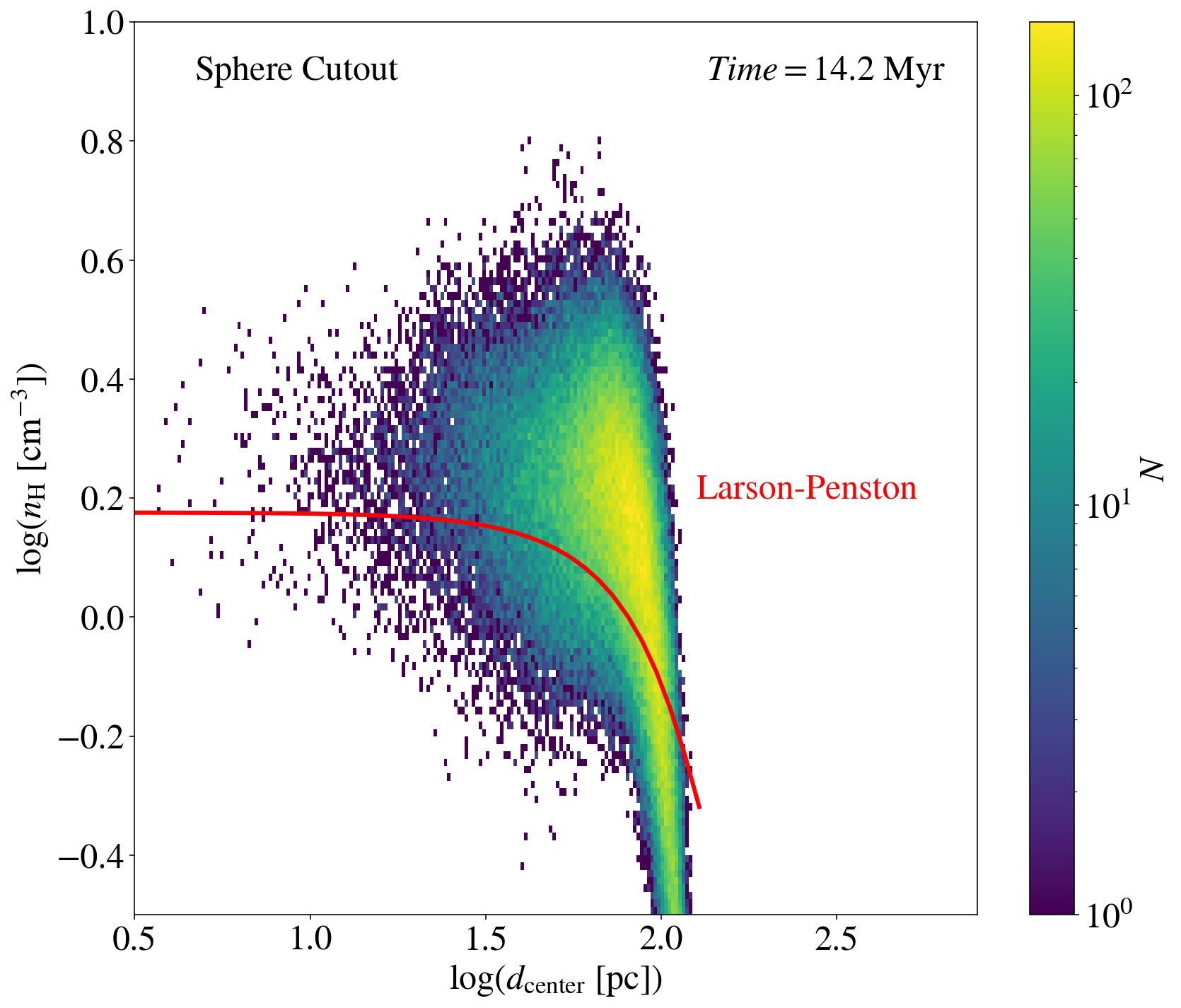}{0.5\textwidth}{}
}
\vspace{-3em}
\gridline{
\fig{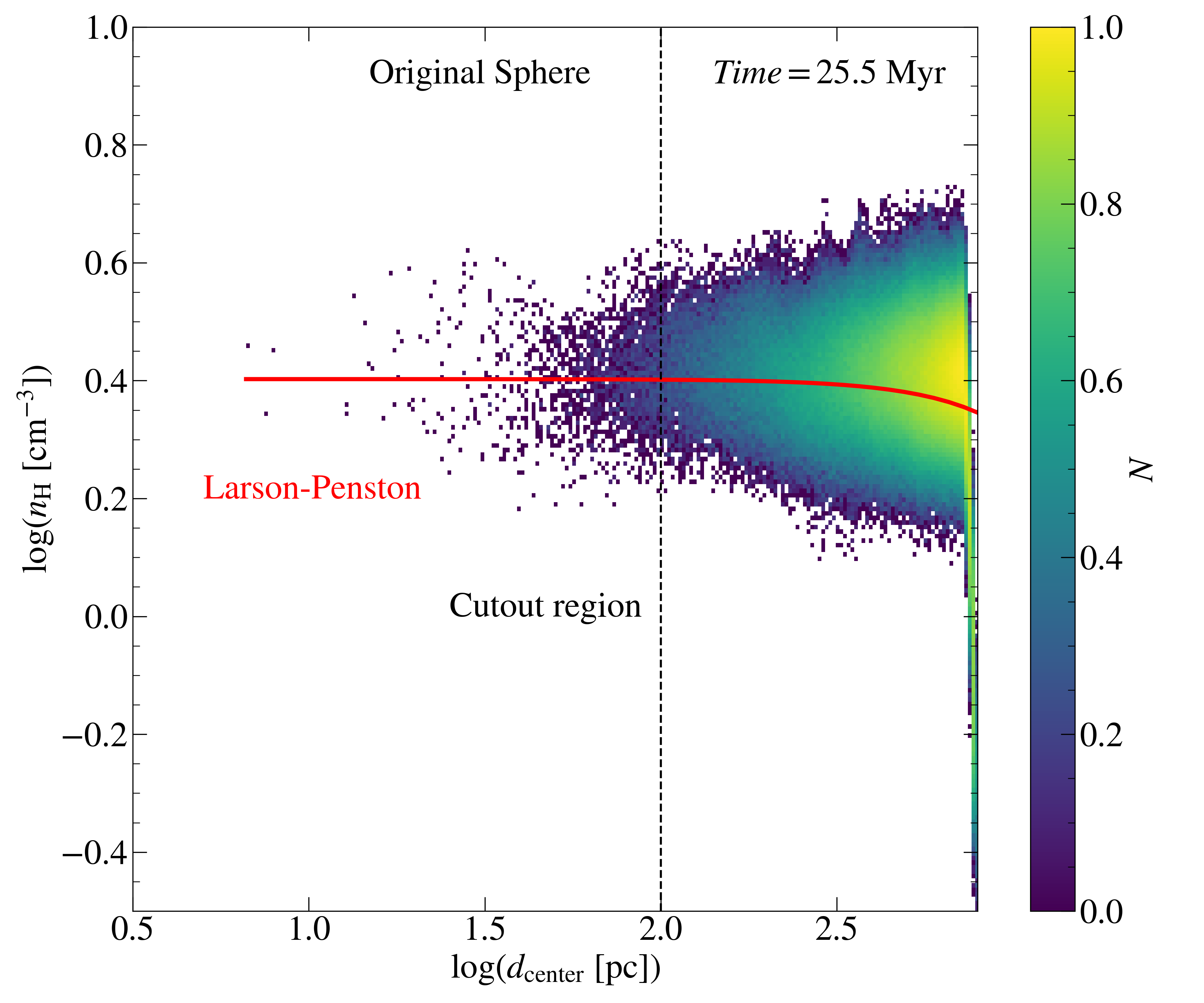}{0.5\textwidth}{}
\fig{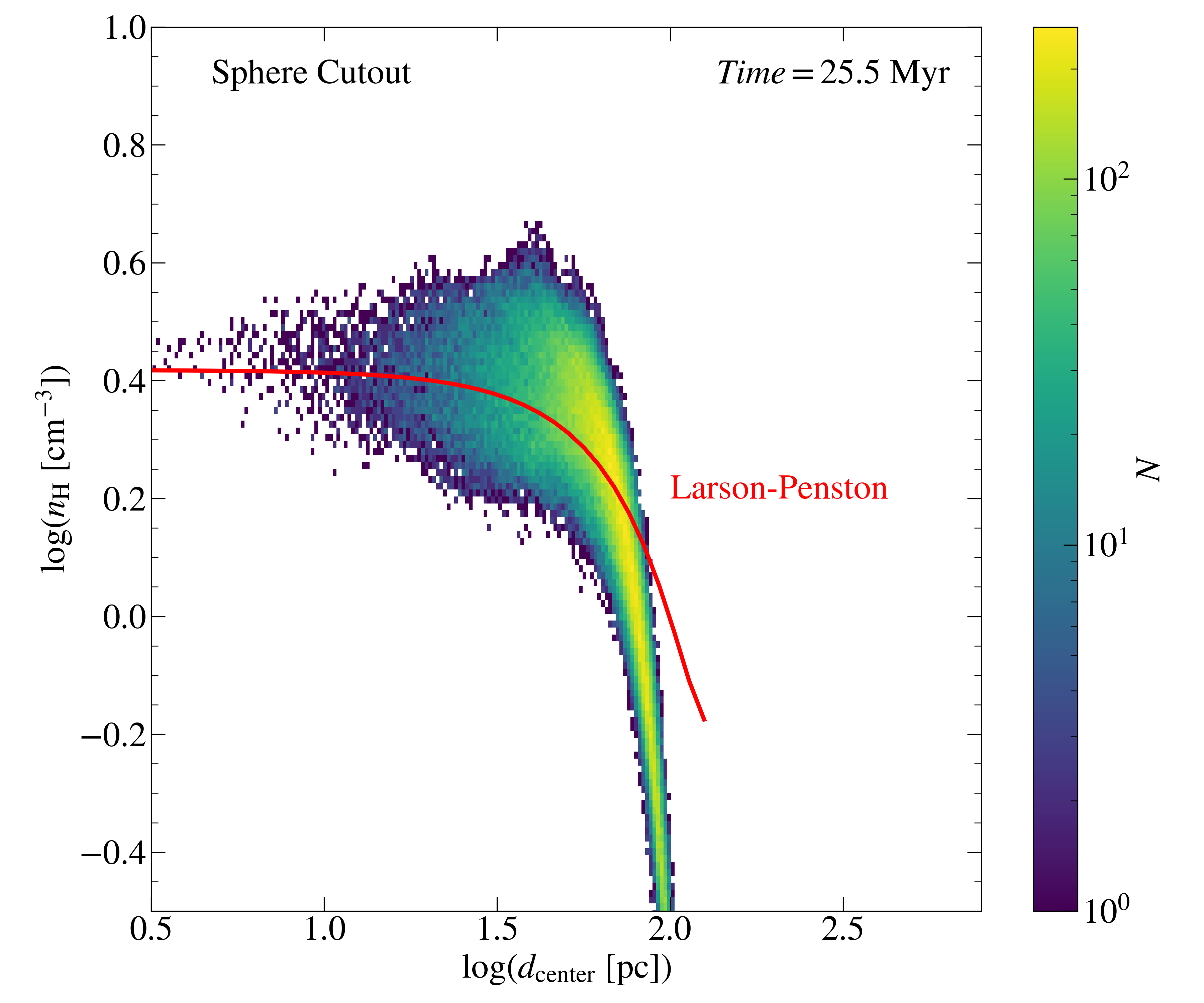}{0.5\textwidth}{}
}
\vspace{-3em}
\caption{Gas hydrogen number density vs. distance from the center for an idealized gas sphere and its cutout at two different times. The sphere was run for $\sim13$ Myr and its inner 100 pc was converted into a cutout with a higher resolution with the method described in Sec.~\ref{sec:simset}. Both the cutout and the sphere were then evolved to $25.5$~Myr. While the gas density at the outer edge for both the sphere and cutout artificially drops due to the vacuum boundary conditions, the gaseous collapse near the center of the sphere is better resolved in the cutout. The evolution of the inner region can be well fit with the analytical Larson-Penston solution of isothermal collapse \citep{Larson1969,Penston1969}, where the core rises in density with time. The original sphere with lower resolution could not resolve the gas collapse on small scales, while the cutout can trace the collapse to smaller scales.}
    \label{fig:appsphere}
\end{figure*}

To confirm the validity of the cutout method, we ran a simulation of an idealized gas sphere. The sphere is initialized with a uniform gas density and total mass of $1.5\times10^8$ M$_\odot$, radius 1,000 pc, and gas particle mass 150 M$_\odot$. We evolve the gas sphere without SF effects, similar to the cutout runs, thus focusing on the gas collapse.

We run the gas sphere for $\sim13$ Myr and apply a similar method as described in Sec.~\ref{sec:simset} to prepare a cutout of a 100 pc region inside the gas sphere. The particles within 100 pc of the sphere center are split by a factor of 100, resulting in 1.5 M$_\odot$ particles. Both the cutout and the original sphere are then run until $\sim25.5$~Myr to compare the gas collapse for the two resolutions. We find that, as expected, the cutout is able to resolve the gas collapse to smaller scales than the original sphere, as shown in Fig.~\ref{fig:appsphere}. The gas expands at the outer edge of the simulations due to the vacuum boundary conditions, but the collapse closer to the center of the sphere is better resolved in the cutout run. We further fit a functional form of the analytical Larson-Penston model for gravitational collapse of an isothermal sphere \citep{Larson1969,Penston1969} at two different times, finding good agreement in the inner regions, with the expected behavior of increasing central densities over time. 
The cutout approach can thus be used to better resolve the small-scale hydrodynamic physics at a limited computational cost, offering an efficient technique to the broader numerical toolkit.







\bibliography{ms}{}
\bibliographystyle{aasjournal}



\end{document}